\documentclass[%
 reprint,
superscriptaddress,
 amsmath,amssymb,
 aps,
onecolumn
]{revtex4-2}

\usepackage[utf8]{inputenc}
\usepackage{xcolor} 
\usepackage{braket}
\usepackage{physics}
\usepackage{comment}
\usepackage{mathrsfs}
\usepackage{enumerate}

\usepackage[breaklinks=true]{hyperref}
\usepackage{here}

\usepackage{footmisc}

\usepackage{mathtools} 
\mathtoolsset{showonlyrefs,showmanualtags}
\usepackage{graphicx}

\newcommand{\ms}[1]{\mathscr{#1}}
\newcommand{\mcl}[1]{\mathcal{#1}}

\newcommand{\prob}{\mathrm{prob}}
\newcommand{\opt}{\mathrm{opt}}

\date{\today}
\usepackage{amsthm}
\theoremstyle{definition}

\begin{document}
\title{Optimal Control in Nearly-Adiabatic Two-Level Quantum Systems via Time-Dependent Resonance}

\author{Takayuki Suzuki}
\affiliation{Quantum ICT Laboratory, National Institute of Information and Communications Technology, 4-2-1 Nukui-Kitamachi, Koganei,
Tokyo 184-8795, Japan}
\affiliation{TRIP Headquarters, RIKEN, 2-1 Hirosawa, Wako, Saitama 351-0198, Japan}

\begin{abstract}
In this study, we theoretically analyzed a control protocol based on ``time-dependent resonance" in nearly adiabatic two-level quantum systems, demonstrating that it exhibits properties equivalent to adiabatic control. This protocol is based on ``time-dependent resonance", where the frequency corresponds to the time-dependent energy gap. Through numerical calculations, we showed that this protocol serves as an optimal control protocol. This approach enables efficient and high-precision transitions to the target state. Our findings provide a new perspective on quantum optimal control theory and suggest potential applications in qubit controls and quantum information processing. 
\end{abstract}

\maketitle

\section{Introduction}\label{Sec:Introduction}
Quantum control has been widely applied in various fields, including quantum computing and quantum sensing, and is recognized as an essential foundation for these technologies~\cite{werschnik2007quantum,geng2016experimental,rembold2020introduction,ansel2024introduction}. Optimal control theory plays a critical role in the design of control protocols that minimize specific costs, and its applications have extended to quantum annealing~\cite{kadowaki1998quantum,farhi2001quantum} and the Quantum Approximate Optimization Algorithm (QAOA)~\cite{farhi2014quantum} which is a type of variational quantum algorithm~\cite{yang2017optimizing,brady2021optimal}. One advantage of this method is that, unlike shortcut to adiabaticity~\cite{guery2019shortcuts}, it does not require additional terms in the Hamiltonian, making it both simpler and more practical for experimental implementations. However, applying optimal control theory to high-dimensional systems becomes often impractical due to the increase in computational complexity. Therefore, understanding the analytical behavior or characteristic properties of optimal control remains a significant challenge in advancing quantum control theory~\cite{caneva2009optimal}.

Adiabatic control, as one of the quantum control methods, is applied in various experiments~\cite{hu2019adiabatic,kandel2021adiabatic,kral2007colloquium}. Furthermore, in the context of quantum algorithms, it plays a significant role in several algorithms, including the adiabatic state preparation algorithm~\cite{costa2022optimal,kovalsky2023self,xu2017experimental,albash2018adiabatic,wen2019experimental}. Ideally, the system should be manipulated infinitely slowly, but in practice, it is operated at a finite speed, which introduces errors. We refer to the regime where these errors remain small as the nearly-adiabatic regime.
Previous study has shown that optimal control in the nearly-adiabatic regime involves characteristic oscillations where the driving frequency is determined by the time-dependent energy gap of the Hamiltonian~\cite{brady2021behavior}. 
This finding also suggests that optimal control in the adiabatic regime could potentially be determined analytically. To address this issue, we applied optimal control to the simplest yet nontrivial case of a two-level system. By approximately solving the dynamics of this system, we demonstrated that ``time-dependent resonance" plays an important role in controlling the dynamics, enabling the probability of remaining in the ground state to approach $1$ with a minimal cost.

Furthermore, by numerically solving for the optimal control that minimizes the transition probability to the excited state as the cost function, we confirmed that the resulting optimal control is consistent with the control protocol based on ``time-dependent resonance". This study provides new insights into quantum optimal control in the nearly-adiabatic regime and offers valuable knowledge that contributes to the advancement of quantum control theory and its potential applications in quantum technologies.

\section{Analysis of the Dynamics}\label{Sec:Analysis}

In this section, we analyze the dynamics under the following ``time-dependent resonance" Hamiltonian:
\begin{align}
    H_n(t)&=\qty(v^{n+1}t^n+\sum_{k=0}^{n-1} A_{n,k}(t)\sin \phi_n(t,t_{r,k}))\sigma_z+\Delta \sigma_x,
\end{align}
where
\begin{align}
    A_{n,k}(t)&=\frac{-\alpha_k}{\mcl{E}_n  (t)},\quad 
    \mcl{E}_n(t)=\sqrt{v^{2n+2}t^{2n}+\Delta^2},\\
    \phi_n(t,t_0)&= 2\int_{t_0}^{t} \mcl{E}_n  (s)ds ,\label{eq:int_E}\\
\end{align}
and $n$ is assumed to be odd for simplicity. $t_{r,k}$ represents the time when the Stokes line intersects the real axis, which is explained later. Here are some comments on this Hamiltonian. First, $\mcl{E}_n(t)$ is the positive eigenvalue of the Hamiltonian when there are no oscillations (i.e., $\alpha_k=0$):
\begin{align}
    \mcl{H}_n(t)&=v^{n+1}t^n\sigma_z+\Delta \sigma_x.
\end{align}
Since the frequency in the Hamiltonian $H_n(t)$ is given by $\dot\phi_n(t,t_0)=2\mcl{E}_n(t)$, it can be understood as a resonance oscillation corresponding to the time-dependent energy gap. For this reason, this Hamiltonian is referred to as ``time-dependent resonant" Hamiltonian. 
We note that previous study has considered oscillations of the form $\sin(2\mcl{E}_n(t)t)$~\cite{brady2021behavior}.

Further definitions of notations are provided. The eigenstates of the Hamiltonian $H_n(t)$ are denoted by $\ket{E_{n,\pm}(t)}$, and the time-evolution operator is expressed as $U_n(t,t_0)$. On the other hand, the eigenstates of the Hamiltonian $\mcl{H}_n(t)$ are denoted by $\ket{\mcl{E}_{n,\pm}(t)}$, and the time-evolution operator is expressed as $\mathcal{U}_n(t,t_0)$. In particular, the eigenstates of this Hamiltonian $\mcl{H}_n(t)$ are expressed as
\begin{align}
    \ket{\mcl{E}_{n,+}(t)}&=\mqty(\cos\frac{\theta_n (t)}{2}\\\sin\frac{\theta_n(t)}{2}),\quad \ket{\mcl{E}_{n,-}(t)}=\mqty(\sin\frac{\theta_n(t)}{2}\\-\cos\frac{\theta_n (t)}{2}),\\
    \tan\theta_n(t)&=\frac{\Delta}{v^{n+1}t}.
\end{align}
From now on, the initial time $t_0$ is omitted in the notation of the time-evolution operator for simplicity. Furthermore, by introducing the dimensionless quantity $\tau = vt$, the Schr\"odinger equation can be rewritten as
\begin{align}
    &i\frac{d}{d\tau}U_n(\tau)\\
    &=\qty(\qty(\tau^n+\sum_{k=0}^{n-1}\tilde A_{n,k}(\tau)\sin \phi_n(\tau,\tau_{r,k}))\sigma_z+\tilde \Delta \sigma_x)U_n(\tau),
\end{align}
where the Hamiltonian is also expressed in a dimensionless form with $\tilde \circ = \circ / v$. 
Under this dimensionless Hamiltonian, we analyze how the transition probability between eigenstates behaves. Let the initial time be $\tau_0$ and the final time be $\tau_f$. The transition probability between eigenstates is given by
\begin{align}
    P_e&:=\qty|\bra{E_{n,+}(\tau_f)}U_n(\tau_f)\ket{E_{n,-}(\tau_0)}|^2.
\end{align}

\begin{figure}[h]
    \centering
    \includegraphics[width=0.43\linewidth]{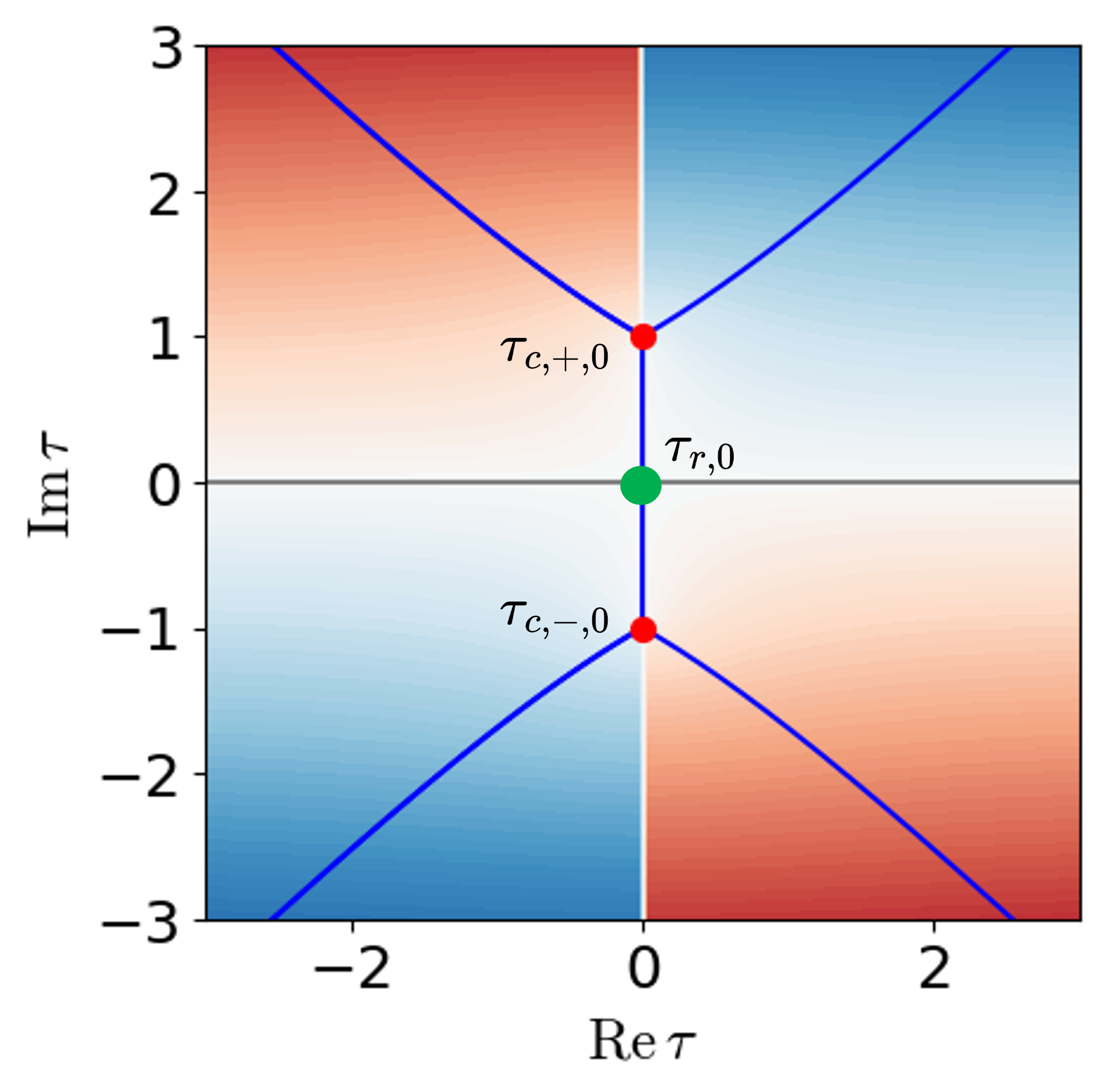}
    \includegraphics[width=0.43\linewidth]{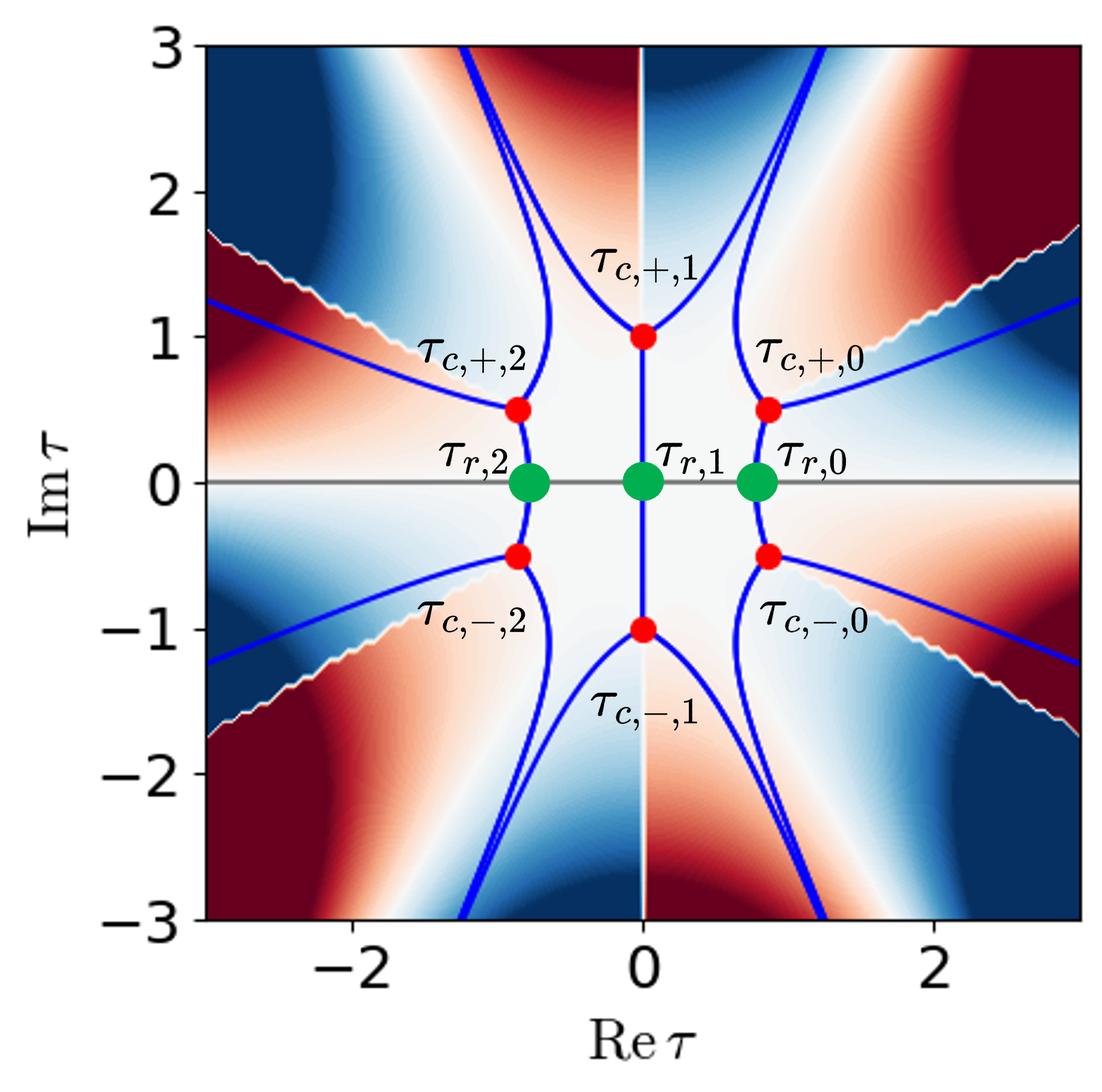}
    \caption{Stokes diagrams for $n=1$ and $n=3$. The red dots represent the turning points $\tau_{c, \pm, k}$, and the blue lines represent the Stokes lines. The times when the Stokes lines originating from $\tau_{c, +, k}$ intersect the real axis are denoted as $\tau_{r,k}$ and are shown as green dots.}
    \label{fig:stokes_point}
\end{figure}

First, we consider the dynamics under a Hamiltonian without oscillatory terms:
\begin{align}
    \mcl{H}_n(\tau)=\tau^n\sigma_z+\tilde \Delta\sigma_x.
\end{align}
For $n=1$, this is known as the Landau-Zener-St\"uckelberg-Majorana (LZSM) model~\cite{landau1932theorie,zener1932non,stueckelberg1932two,majorana1932atoms}, which can be solved exactly. However, for other values of $n$, the solution is not known. However, in the adiabatic regime, the transition amplitude over an infinite time interval has been studied~\cite{vitanov1999nonlinear,lehto2012superparabolic} using the Dykhne--Davis--Pechukas (DDP) formula \cite{dykhne1962adiabatic,davis1976nonadiabatic}.
To analyze the dynamics in the adiabatic regime, the turning points, which are the zeros of the energy $\mcl{E}_n(\tau)$, and the Stokes lines originating from these turning points play an important role~\cite{suppl_ref}. 
The definition of the Stokes line is as follows:
\begin{align}
    \Re \int_{\tau_c}^\tau  \mcl{E}_n(s) ds= 0.
\end{align}
For the Hamiltonian $\mcl{H}_n(t)$,
\begin{align}
    \tau_{c, \pm, k}=e^{ \pm i\left(\frac{\pi}{2 n}+\frac{k}{n} \pi\right)}\tilde{\Delta}^{\frac{1}{n}}, \quad k=0, \ldots, n-1
\end{align}
are the turning points. Using this, the transition amplitude for $\tau_0\simeq -\infty$ and $\tau_f\simeq \infty$ can be approximated as follows~\cite{suppl_ref}:
\begin{widetext}
\begin{align}
    \bra{\mcl{E}_{n,+}(\tau_f)}\mathcal{U}_n(\tau_f)\ket{\mcl{E}_{n,-}(\tau_0)}=\sum_{k=0}^{n-1}(-1)^ke^{i \int_{\tau_0}^{\tau_{r,k}}\mcl{E}_n(s)ds}e^{-i \int_{\tau_{r,k}}^{\tau_f} \mcl{E}_n(s) d s}e^{-2\Im\int_{\tau_{r,k}}^{\tau_{c,+,k}}\mcl{E}_n(s)ds},\label{eq:free_unitary}
\end{align}
\end{widetext}
where $\tau_{r,k}$ represents the time when the Stokes line originating from $\tau_{c,+,k}$ intersects the real axis (Fig.~\ref{fig:stokes_point}). It is noted that we can understand from this expression that the exact WKB analysis corresponds to the adiabatic-impulse approximation~\cite{shevchenko2010landau,suzuki2022generalized} in the nearly-adiabatic region.

Next, we consider the dynamics under the Hamiltonian $H_n(\tau)$, which includes oscillatory terms. Treating the Hamiltonian $\mcl{H}_n(\tau)$, without oscillatory terms, as the free Hamiltonian, the time-evolution operator can be approximated perturbatively as
\begin{align}
    U_n(\tau)
    &\simeq \mathcal{U}_n  (\tau)-i\mathcal{U}_n  (\tau)\int_{\tau_0}^\tau  F(s)ds,\\
    F(s)&=\mathcal{U}_n^\dagger(s) \sigma_z  \mathcal{U}_n  (s)\sum_{k=0}^{n-1} \tilde A_{n,k}(s)\sin \phi_n(s,\tau_{r,k}),
\end{align}
where the picture is called as ``Furry picture"~\cite{furry1951bound}. Thus, 
the transition probability can be approximated as
\begin{align}
    P_e&\simeq \biggl|\bra{\mcl{E}_{n,+}(\tau_f)}\mathcal{U}_n  (\tau_f)\ket{\mcl{E}_{n,-}(\tau_0)}\\
    &-i\bra{\mcl{E}_{n,+}(\tau_f)}\mathcal{U}_n  (\tau_f)\int_{\tau_0}^{\tau_f}  F(s)ds\ket{\mcl{E}_{n,-}(\tau_0)}\biggr|^2,\label{eq:furry_case_I}
\end{align}
where it was assumed that the amplitude is sufficiently small at $\tau=\tau_0$ and $\tau_f$, allowing the approximation $\ket{E_{n,\pm}(\tau)}\simeq \ket{\mcl{E}_{n,\pm}(\tau)}$.

The first term in \eqref{eq:furry_case_I} corresponds to the free dynamics~\eqref{eq:free_unitary}. On the other hand, the second term represents the first-order perturbative term arising from the oscillations. This term can be approximated as
\begin{align}
    &-i\bra{\mcl{E}_{n,+}(\tau_f)}\mathcal{U}_n (\tau_f)\int_{\tau_0}^{\tau_f}  F(s)ds\ket{\mcl{E}_{n,-}(\tau_0)}\\
    &\simeq  -\frac{1}{2}e^{-i\int_{\tau_0}^{\tau_f}\mcl{E}_n(s)ds} \sum_{k=0}^{n-1} e^{i\phi_n(\tau_{r,k},\tau_0)}\int_{\tau_0}^{\tau_f}\frac{\tilde{\alpha}_k\tilde\Delta }{\mcl{E}^2_n  (s)}ds \label{eq:phase_destruct},
\end{align}
where the DDP formula was applied to ignore the counter-rotating term, and terms negligible in the adiabatic regime were omitted.
Thus, the transition probability can be approximated as
\begin{align}
    P_e&\simeq \biggl| \sum_{k=0}^{n-1} e^{i \int_{\tau_0}^{\tau_{r,k}}\mcl{E}_n(s)ds}e^{-i \int_{\tau_{r,k}}^{\tau_f} \mcl{E}_n(s) d s}\\
    &\times \qty((-1)^ke^{-2\Im\int_{\tau_{r,k}}^{\tau_{c,+,k}}\mcl{E}_n(s)ds}-\frac{\tilde{\alpha}_k\tilde\Delta }{2}\int_{\tau_0}^{\tau_f}\frac{1}{\mcl{E}^2_n  (s)}ds)\biggr|^2,\label{eq:trans_prob_case_I}
\end{align}
and the condition for the transition probability to be zero is given by
\begin{align}
    \tilde \alpha_k&=\frac{2(-1)^ke^{-2\Im\int_{\tau_{r,k}}^{\tau_{c,+,k}}\mcl{E}_n(s)ds}}{\tilde\Delta   \int_{\tau_0}^{\tau_f}\frac{1}{\mcl{E}^2_n  (s)}ds}=:\tilde\alpha_{k,\opt}.
\end{align}

\begin{figure}[h]
    \centering
    \includegraphics[width=0.45\linewidth]{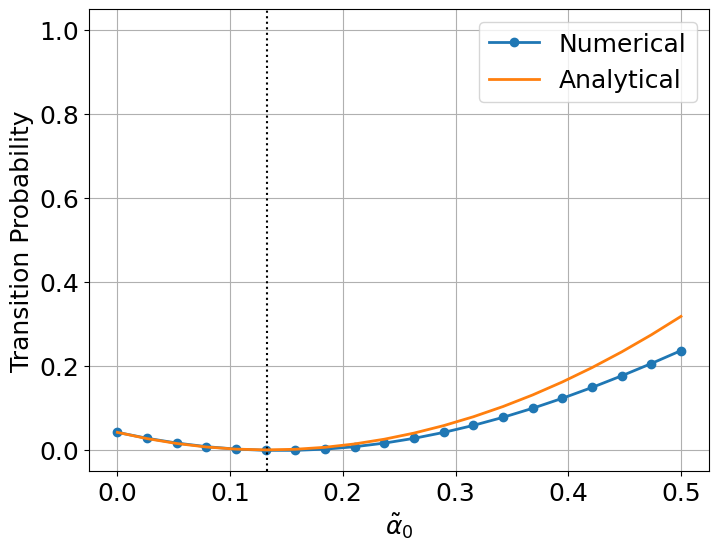}
    \includegraphics[width=0.45\linewidth]{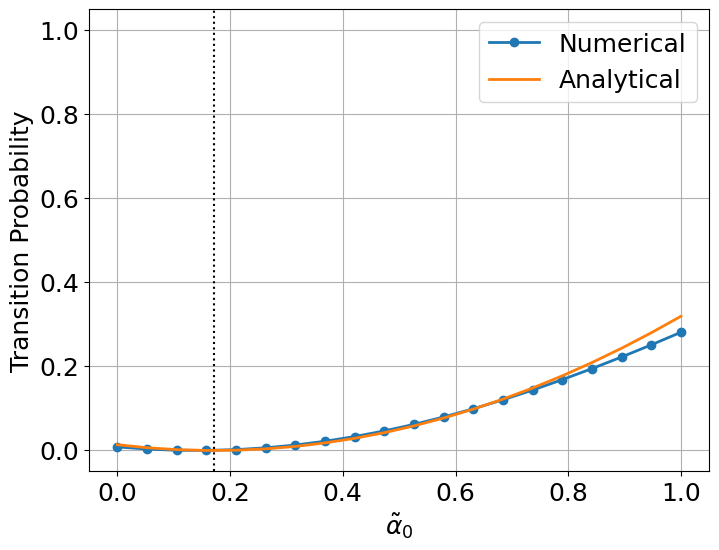}
    \caption{Comparison between the numerical results and the analytical approximation of $P_e$ for (Left) $n=1, \tau_f=-\tau_0=100$ and $\tilde \Delta = 1$ and (Right) $n=3, \tau_f=-\tau_0=5, \tilde \Delta = 1$. It can be seen that the results agree well in the region of small amplitudes. The black dashed line indicates $\tilde\alpha_{0,\opt}$. In the case of $n=3$, the optimal values for $\tilde{\alpha}_1$ and $\tilde{\alpha}_2$ were used.}
    \label{fig:numerical_Delta}
\end{figure}

Let us compute using a specific Hamiltonian for the case of $n=1$ and $n=3$. 
In the long-time limit $-\tau_0\simeq \tau_f\simeq \infty$, 
the optimal amplitude for $n=1$ is
\begin{align}
    \tilde\alpha_{0,\opt}= \frac{2}{\pi}e^{-\tilde\Delta^2\pi/2}\label{eq:optimal_alpha_n_1},
\end{align}
and for $n=3$ is
\begin{align}
    \tilde\alpha_{k,\opt}&=\frac{3(-1)^ke^{-\tilde\Delta^{4/3} \sin\qty(\pi\qty(\frac{1}{6}+\frac{k}{3}) )\frac{\sqrt{\pi} \Gamma\left(\frac{7}{6}\right)}{ \Gamma\left(\frac{5}{3}\right)}}}{\tilde\Delta^{-2/3}  \pi},\ k=0,1,2.\label{eq:optimal_alpha_n_3}
\end{align}
A comparison between the analytical results \eqref{eq:trans_prob_case_I} and numerical calculations of the transition probability is shown in Fig.~\ref{fig:numerical_Delta}. It is observed that the results agree well in the region of small amplitude.

Now, let us comment on these results. First, it is evident that there exists an amplitude $\tilde\alpha_0$ in the nearly-adiabatic regime and long-time limit, where the derived approximation is valid, that makes the transition probability $0$.
This effect results from the mitigation of transitions caused by the minimum energy gap at $\tau=0$ in the absence of oscillations. Furthermore, for $0<\tilde\alpha_k<\tilde\alpha_{k,\opt}$, $\partial P_e(\tilde{\boldsymbol{\alpha}})/d\tilde\alpha_k<0$ holds, indicating that the transition probability decreases monotonically from $\tilde\alpha_k=0$. Therefore, when numerically determining the optimal control, this control may be selected without encountering any barriers.

\begin{figure}[h]
    \centering
    \includegraphics[width=0.51\linewidth]{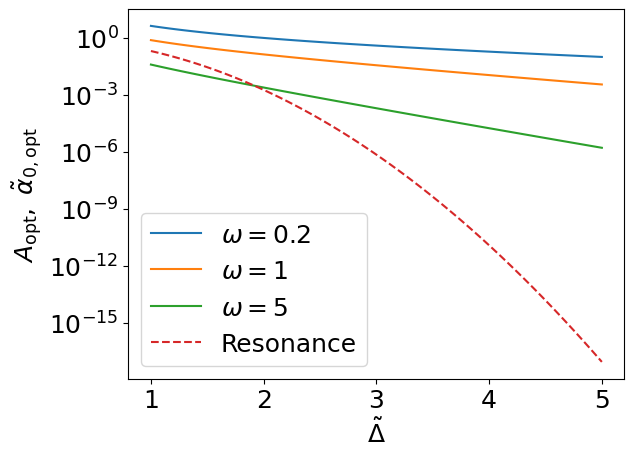}
    \caption{Comparison of optimal amplitudes for resonance and harmonic oscillations. It is found that the more adiabatic the process becomes, the ``time-dependent resonance" can suppress the transition probability with a smaller amplitude.}
    \label{fig:compare_1f1}
\end{figure}

Next, we focus on the results for $n=1$. It is found that the optimal amplitude $\tilde\alpha_{0,\opt}$ decreases exponentially with $\tilde\Delta$. This behavior results from the ``time-dependent resonant" term \eqref{eq:int_E}. Due to this term, the cancellation of the dynamical phase from adiabatic time evolution and the oscillation phase happens in \eqref{eq:phase_destruct}. Let us consider the case of harmonic oscillations with a constant amplitude $A$ and phase $\tilde\omega \tau$ for cases where the oscillation phase and dynamical phase do not cancel. The transition probability in this case becomes~\cite{suzuki2023kibble}
\begin{align}
    &\qty|\bra{E_+(\tau_f)}U(\tau_f)\ket{E_-(\tau_0)}|^2 \\
    &\simeq e^{-\pi\tilde\Delta^2}\qty|1+ \sqrt{2}C A \pi {}_1\tilde F_1\qty(-i\frac{\tilde\Delta^2}{2},0,i\frac{\tilde\omega}{2})|^2,
\end{align}
where $C$ is a suitable constant satisfying $|C|=1$. 
It is necessary to set at least $A_{\opt}^{-1}\propto {}_1\tilde F_1\qty(-i\frac{\tilde\Delta^2}{2},0,i\frac{\tilde\omega}{2})$ to make transition probability $0$. A comparison between the optimal amplitudes for these cases and ``time-dependent resonant" case is shown in Fig.~\ref{fig:compare_1f1}, which demonstrates that ``time-dependent resonance" achieves the desired transitions with smaller amplitudes in nearly-adiabatic regime.

Finally, it is important to note the aspects that also influence cases where the goal is to enhance transition probabilities, such as in the dynamically assisted Schwinger mechanism~\cite{taya2020dynamically}. In the case of ``time-dependent resonance", for amplitudes sufficiently large compared to the LZSM transition probability $e^{-\tilde\Delta^2\pi/2}$, the transition probability scales proportionally to the square of the amplitude. In contrast, for harmonic oscillations, the transition amplitude scales approximately as $A^2 e^{-\pi\tilde\Delta^2}\qty|{}_1\tilde F_1\qty(-i\frac{\tilde\Delta^2}{2},0,i\frac{\tilde\omega}{2})|^2$, which decreases exponentially with $\tilde\Delta$, as shown in Fig.~\ref{fig:compare_1f1}. Thus, to achieve an $O(1)$ transition amplitude, the amplitude $A$ must increase exponentially with $\tilde \Delta$. Based on these discussions, it is evident that ``time-dependent resonance" provide an efficient mechanism for enabling state operations in the nearly-adiabatic regime.

From the above, it was found that ``time-dependent resonant" Hamiltonian induces state transitions with small amplitudes in the nearly-adiabatic regime. Next, we numerically demonstrate that ``time-dependent resonant" Hamiltonian provides optimal control for state transitions.

\section{Comparison with optimal control}\label{Sec:Optimal_control}

We consider a two-level system Hamiltonian as follows:
\begin{align}
    H(t)&=u(t)\sigma_z+\Delta \sigma_x, \quad t\in[-T,T],\ u(t)\in [u_-,u_+].
\end{align}
We set the initial state as the ground state of $u_- \sigma_z+\Delta\sigma_x$, and the goal is to obtain the ground state of $H_C := u_+ \sigma_z+\Delta \sigma_x$ at the final time $T$. The problem of finding the control $u(t)$ that minimizes the expectation value of $H_C$ at the final time can be formulated as a minimization of the following functional \cite{brady2021optimal}:
\begin{align}
&J[\ket{x(t)},\bra{k(t)},u(t)]=  \bra{x(T)}H_C\ket{x(T)} \\
&+\int_{-T}^{T} d t\qty(\bra{k(t)}\qty[-\frac{d}{d t}-iH(t)]\ket{x(t)}+c.c.),\label{eq:functional}
\end{align}
where the final state $\ket{x(T)}$ is not fixed. The initial control before optimization is denoted as $u_0(t)$ and we assume $u_0(t)=v^{n+1}t^n$. 
We derive the dimensionless quantities in the same manner as discussed previously and use them in the subsequent discussion. More detailed optimal control conditions and experimental results are summarized in ~\cite{suppl_ref}.

\begin{figure}[h]
    \centering
    \includegraphics[width=0.385\linewidth]{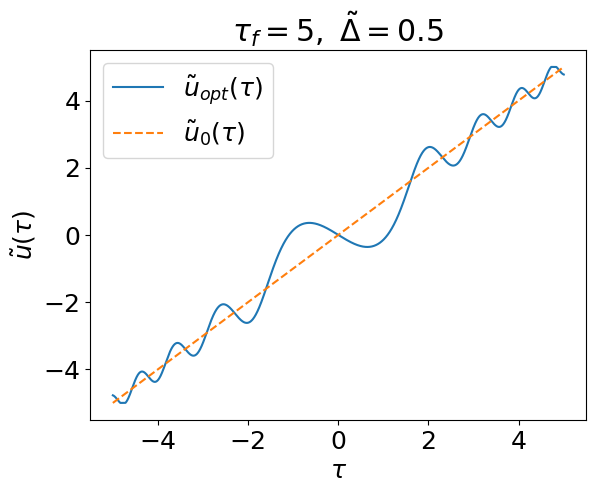}
    \includegraphics[width=0.385\linewidth]{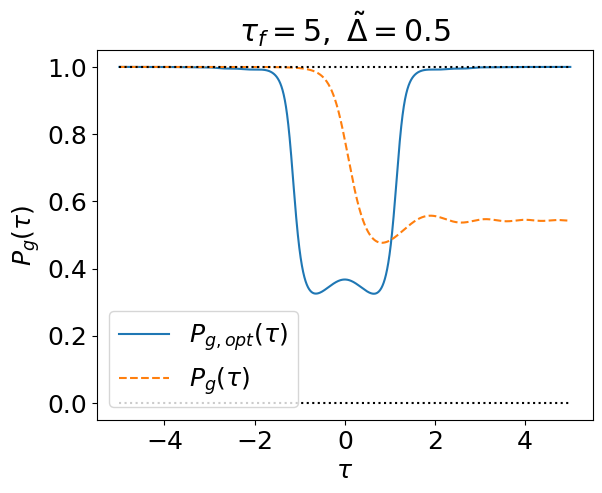}
    \caption{Plots for $\tilde \Delta = 1/2, \tau_f =5, n=1$: (Left) Optimal control $\tilde u_{\opt}(\tau)$ and (Right) probability of obtaining the ground state $P_{g,\opt}(\tau)=|\bra{\psi_{\opt}(\tau)}\ket{E_{1,-}(\tau)}|^2$ under this optimal control. Here, $\ket{\psi_{\opt}(\tau)}$ represents the state evolving under the optimal control, and $\ket{E_{1,-}(\tau)}$ is the instantaneous eigenstate corresponding to the negative eigenvalue of the Hamiltonian $H_1(\tau)$. The dashed lines represent the (Left) initial control and (Right) the probability of obtaining the ground state under the initial control.}
    \label{fig:T5}
\end{figure}

We conduct numerical experiments for the problem. Below, we set $\tilde u_+ = -\tilde u_- = \tau_f^n$.
The results of the optimal control for $\tilde \Delta = 1/2, \tau_f = 5, n=1$ are shown in Fig.~\ref{fig:T5}. As clear from the figure, the optimal control $\tilde u_{\opt}(\tau)$ oscillates around the initial control $\tilde u_0(\tau)$. In particular, the amplitude near $\tau=0$ is the largest and significantly affects the dynamics.

\begin{figure}[h]
    \centering
    \includegraphics[width=0.51\linewidth]{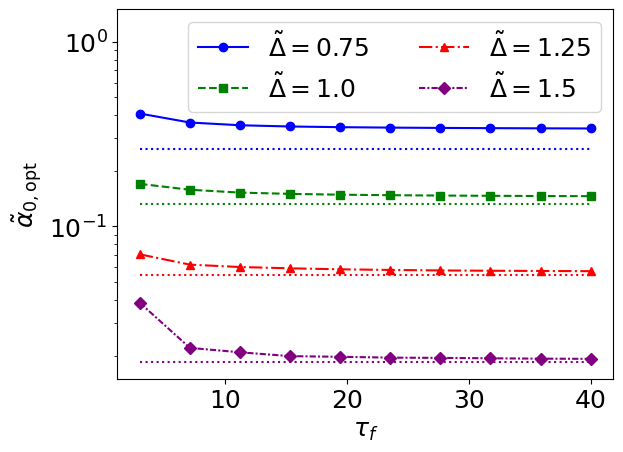}
    \caption{Comparison of (solid line) the optimal amplitude of oscillation obtained from fitting and (dotted line) the analytically derived amplitude $\tilde\alpha_{0,\opt}$ \eqref{eq:optimal_alpha_n_1}. For sufficiently large $\tau_f$ and $\tilde\Delta$ in the nearly-adiabatic region, the numerical and analytical results agree well.}
    \label{fig:tildealpha}
\end{figure}

The optimal and analytically derived amplitudes of oscillation $\tilde\alpha_{\opt}$ for adiabaticity and large time are compared in Fig.~\ref{fig:tildealpha}. In the nearly-adiabatic regime and long-time limit, it is found that the numerically determined optimal control coincides with the ``time-dependent resonant" Hamiltonian. On the other hand, when the long-time limit does not hold, it becomes necessary to revise the derivation. Such a scenario is particularly important in cases involving time dependencies similar to those in quantum annealing, as discussed in previous studies \cite{brady2021behavior,brady2021optimal}. The analysis of cases where the long-time limit does not hold, including through the context of this scenario, is discussed in detail in~\cite{suppl_ref}.

\begin{figure}[h]
    \centering
    \includegraphics[width=0.31\linewidth]{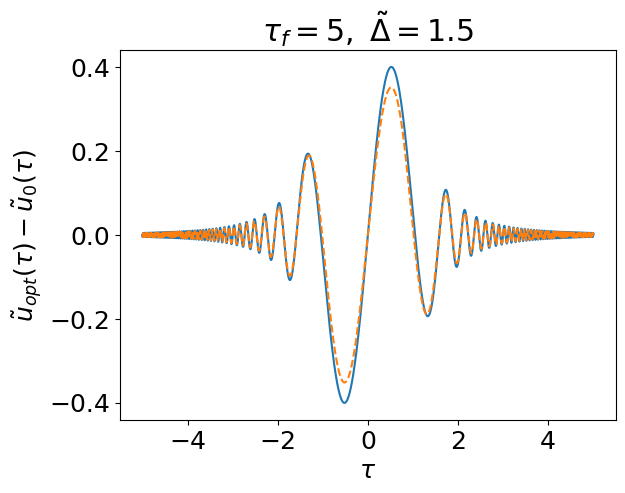}
    \includegraphics[width=0.31\linewidth]{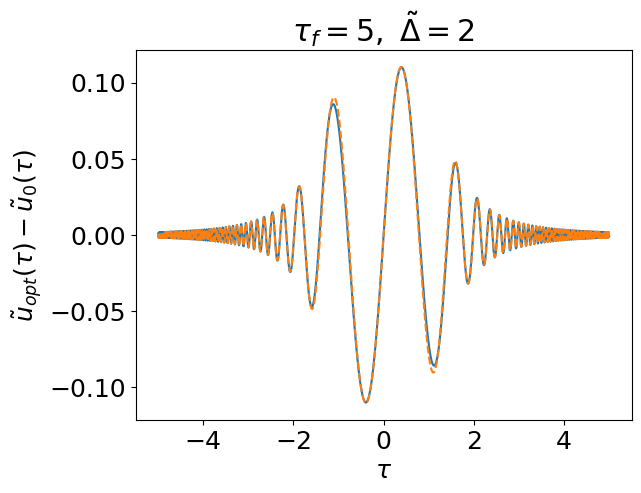}
    \includegraphics[width=0.31\linewidth]{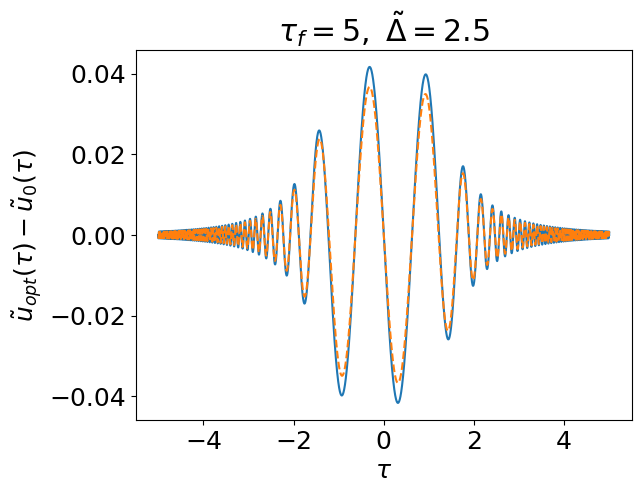}
    \caption{Plots of the difference between the optimal control $\tilde{u}_{\mathrm{opt}}(\tau)$ and the initial control $\tilde{u}_0(\tau)$ for $\tau_f = 5, n = 3$: (left) $\tilde{\Delta} = 3/2$, (center) $\tilde{\Delta} = 2$, and (right) $\tilde{\Delta} = 5/2$. The solid line represents the numerically optimized control, while the dashed line represents the ``time-dependent resonant" control satisfying \eqref{eq:optimal_alpha_n_3}. It can be seen that these controls agree well in the nearly-adiabatic regime.
    }
    \label{fig:diff_u3_Delta}
\end{figure}

The above analysis corresponds to the case where the initial control has $n=1$. Next, we will examine the case of $n=3$. In this case, the analytically derived ``time-dependent resonant" Hamiltonian consisted of a sum of three oscillations. A comparison between this analytical solution and the specific numerical optimal control is shown in~\ref{fig:diff_u3_Delta}. As can be seen from this figure, the analytically derived control and the numerically determined optimal control agree well.

Finally, let us comment on the extension to multi-level systems. While the method using ``time-dependent resonance" is effective for controlling transitions between two states, it may not be suitable for controlling transitions among three or more states. Therefore, if the transition probability to the second or higher excited states is initially zero in free dynamics, the ``time-dependent resonance" may serve as the optimal control. However, if the transition probability to the second or higher excited states is non-zero, while the transition to the first excited state can be minimized, it cannot be considered the optimal control for maximizing the probability of remaining in the ground state~\cite{suppl_ref}.

\section{Summary}\label{Sec:Summary}
In this study, we analyzed control using ``time-dependent resonant" Hamiltonians, which includes an oscillatory term with a time-dependent energy gap as its frequency, applied to a simple yet nontrivial two-level system. Through the analysis of the dynamics, we concluded that ``time-dependent resonance" is effective for state control and confirmed that it is consistent with the results of numerical optimal control calculations. While we considered control to suppress state transitions, we believe that this approach could also play a significant role in cases such as the dynamically assisted Schwinger model, where the cost is defined by the excitation rate of the state~\cite{taya2020dynamically}. 

\section*{Acknowledgment}

This work was supported by Japan’s MEXT Quantum Leap Flagship Program under Grant No. JPMXS0120319794 and by the RIKEN TRIP initiative (RIKEN Quantum).

\appendix

\section{Exact WKB Analysis}\label{AppSec:Exact_WKB_analysis}

In this section, we briefly summarize the results of \cite{suzuki2024exact}. The exact WKB analysis was originally discussed as an analytical method for second-order differential equations. Similarly, it can also be applied to the Schr\"odinger equation of a two-level system, especially in the adiabatic regime. Here, we consider the following form of the Hamiltonian:
\begin{align}
    H(\tau) = \boldsymbol{d}(\tau) \cdot \boldsymbol{\sigma}.
\end{align}
The eigenvalues of this Hamiltonian are expressed as:
\begin{align}
    E_\pm(\tau) = \pm \sqrt{\boldsymbol{d}^2(\tau)},
\end{align}
where we assume $E_+(\tau) > E_-(\tau)\ \forall\tau\in\mathbb{R}$ and no degeneracy occurs. In this method, the turning points, where the energy becomes zero, play a crucial role. For simplicity, we consider the case where there is only one turning point in the upper half-plane:
\begin{align}
    E_+(\tau_c) = 0,
\end{align}
where $\Im \tau_c > 0$. Furthermore, the set satisfying the following equation is called the Stokes line:
\begin{align}
    \Re \int_{\tau_c}^\tau  E_+(s) ds= 0.
\end{align}
Each time the Stokes line is crossed, the solution is considered to change discretely (though exponentially small). The solution of the Schr\"odinger equation is expressed as:
\begin{align}
    |\psi(\tau)\rangle = \sum_{i=\pm} a_i(\tau)\left|E_i(\tau)\right\rangle.
\end{align}
Rewriting it in the following form:
\begin{align}
    \ket{\psi(\tau)} = \binom{a_-(\tau)}{a_+(\tau)}.
\end{align}
Additionally, the following states under the adiabatic approximation are the WKB solutions:
\begin{align}
\ket{\psi_{A, -}\qty(\tau, \tau_0, \eta)} &= \mqty(0\\e^{-i \int_{\tau_0}^\tau\qty(E_-(s)+g_{-}(s)) d s}), \\
\ket{\psi_{A, +}\qty(\tau, \tau_0, \eta)} &= \mqty(e^{-i \int_{\tau_0}^\tau\qty(E_+(s)+g_{+}(s)) d s}\\0).
\end{align}
Now, when crossing the Stokes line counterclockwise where $\ket{\psi_{A, -}\qty(\tau, \tau_c, \eta)}$ dominates, the solution changes as follows:
\begin{align}
    \mqty(\ket{\psi_{A, +}\qty(\tau, \tau_0, \eta)} \\
    \ket{\psi_{A, -}\qty(\tau, \tau_0, \eta)} ) \to 
    \mqty(1&0\\i\cot \frac{\theta(\tau_c)}{2}e^{-i\int_{\tau_0}^{\tau_c}\qty(E_-(s)-E_+(s)+g_-(s)-g_+(s))ds} &1) 
    \mqty(\ket{\psi_{A, +}\qty(\tau, \tau_0, \eta)}\\ 
    \ket{\psi_{A, -}\qty(\tau, \tau_0, \eta)} ).
\end{align}
Conversely, when crossing the Stokes line counterclockwise where $\ket{\psi_{A, +}\qty(\tau, \tau_c, \eta)}$ dominates, the solution changes as follows:
\begin{align}
    \mqty(\ket{\psi_{A, +}\qty(\tau, \tau_0, \eta)} \\
    \ket{\psi_{A, -}\qty(\tau, \tau_0, \eta)} ) \to 
    \mqty(1&i\tan \frac{\theta(\tau_c)}{2}e^{i\int_{\tau_0}^{\tau_c}\qty(E_-(s)-E_+(s)+g_-(s)-g_+(s))ds} \\
    0&1) 
    \mqty(\ket{\psi_{A, +}\qty(\tau, \tau_0, \eta)}\\ 
    \ket{\psi_{A, -}\qty(\tau, \tau_0, \eta)} ).
\end{align}
Using this, for example, in the case of the nonlinear LZSM model, the WKB solution changes as follows:
\begin{align}
    \mqty(\ket{\psi_{A, +}\qty(\tau, \tau_0, \eta)} \\
    \ket{\psi_{A, -}\qty(\tau, \tau_0, \eta)} ) &\to
    \mqty(1&-\sum_{k=0}^{n-1}(-1)^ke^{i\int_{\tau_0}^{\tau_{c,-,k}}\qty(E_-(s)-E_+(s))ds} \\
    \sum_{k=0}^{n-1}(-1)^ke^{-i\int_{\tau_0}^{\tau_{c,+,k}}\qty(E_-(s)-E_+(s))ds}&1) 
    \mqty(\ket{\psi_{A, +}\qty(\tau, \tau_0, \eta)}\\ 
    \ket{\psi_{A, -}\qty(\tau, \tau_0, \eta)} ),\\
    \tau_{c, \pm, m} &= e^{ \pm i\left(\frac{\pi}{2 n}+\frac{m}{n} \pi\right)}\left(\frac{\Delta}{v^{n+1}}\right)^{\frac{1}{n}}, \quad m = 0, \ldots, n-1.
\end{align}
Here,
\begin{align}
    \sum_{k=0}^{n-1}(-1)^ke^{i\int_{\tau_0}^{\tau_{c,-,k}}\qty(E_+(s)-E_-(s))ds}&=\sum_{k=0}^{n-1}(-1)^ke^{-\Im\int_{\tau_{r,k}}^{\tau_{c,-,k}}\qty(E_+(s)-E_-(s))ds}e^{i\int_{\tau_0}^{\tau_{r,k}}\qty(E_+(s)-E_-(s))ds}
\end{align}
leads to 
\begin{align}
    &\ket{\psi_{A, -}\qty(\tau, \tau_0, \eta)} + \sum_{k=0}^{n-1}(-1)^ke^{i\int_{\tau_0}^{\tau_{c,+,k}}\qty(E_+(s)-E_-(s))ds}\ket{\psi_{A, +}\qty(\tau, \tau_0, \eta)} \\
    &=e^{-i \int_{\tau_0}^\tau E_-(s) d s}\ket{E_-(\tau)} + \sum_{k=0}^{n-1}(-1)^ke^{-i \int_{\tau_0}^{\tau_{r,k}}E_-(s)ds}e^{-i \int_{\tau_{r,k}}^\tau E_+(s) d s}e^{2\Im\int_{\tau_{r,k}}^{\tau_{c,+,k}}E_-(s)ds}\ket{E_+(\tau)} .
\end{align}
It should be noted that this can be regarded as an adiabatic-impulse approximation in the nearly-adiabatic regime~\cite{shevchenko2010landau, suzuki2022generalized}.
Typically, in the adiabatic-impulse approximation, the time of non-adiabatic transitions is determined as the moment where the energy gap becomes $0$ when the off-diagonal component becomes zero. However, in the nearly-adiabatic regime, it is found that even moments without energy level crossing can be considered as part of the impulse region.

\section{Detailed Information of Numerical Experiment for Optimal Control}\label{SubSec:Numerical_experiment}

\subsection{Setting}

The condition for the variation of the functional~\eqref{eq:functional} to be zero is given by
\begin{align}
    i\frac{d}{dt}\ket{x(t)}&=H(t)\ket{x(t)},\label{eq:shore_x}\\
    i\frac{d}{dt}\ket{k(t)}&=H(t)\ket{k(t)},\label{eq:shore_k}\\
    \ket{k(T)}&=H_C\ket{x(T)},\label{eq:boundary_T}\\
    0&=-i\bra{k(t)}\sigma_z\ket{x(t)}+c.c.,\label{eq:exp_value}
\end{align}
From Eq.~\eqref{eq:shore_x}, $\ket{x(t)}$ represents the time-evolved state from the specified initial state. On the other hand, from Eqs.~\eqref{eq:shore_k} and \eqref{eq:boundary_T}, $\ket{k(t)}$ is a (non-normalized) state with a boundary condition defined at the final time. Using these states, the problem is to optimize the control $u(t)$ to satisfy Eq.~\eqref{eq:exp_value}.

The dimensionless equations are as follows:
\begin{align}
    i\frac{d}{d\tau }\ket{\psi(\tau)}&=\qty(\tilde u(\tau)\sigma_z+\tilde \Delta\sigma_x)\ket{\psi(\tau)},
\end{align}
and
\begin{align}
    \tilde u_0(\tau)&=\tau^n,\quad \tilde u(\tau)\in \qty[\tilde u_-,\tilde u_+],\\
    \tau&\in \qty[-vT, vT]=:\qty[\tau_0,\tau_f].
\end{align}

\subsection{Results of numerical calculation}

\begin{figure}[H]
    \centering
    \includegraphics[width=0.31\linewidth]{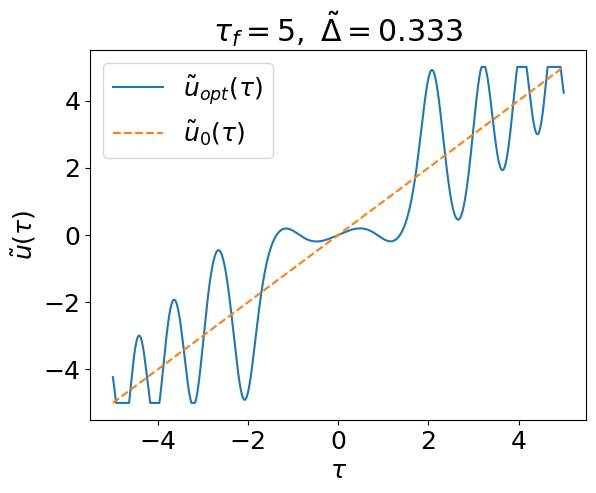}
    \includegraphics[width=0.31\linewidth]{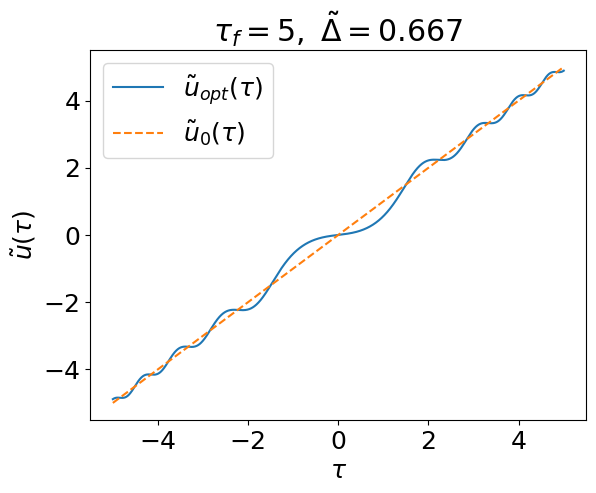}
    \includegraphics[width=0.31\linewidth]{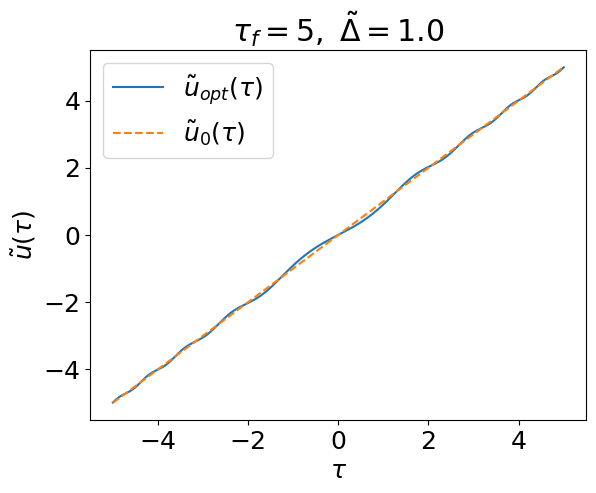}
    \caption{Plots of the optimal control for $\tau_f =5, n=1$: (Left) $\tilde \Delta = 1/3$, (Center) $\tilde \Delta = 2/3$, (Right) $\tilde \Delta = 1$. As $\tilde\Delta$ increases and approaches the adiabatic limit, the oscillation amplitude of the optimal control decreases.}
    \label{fig:u_T5_Delta}
\end{figure}

Next, we compare the behavior of the optimal control as it approaches the adiabatic limit. The results are shown in Fig.~\ref{fig:u_T5_Delta}. As $\tilde\Delta$ increases and approaches the adiabatic limit, the oscillation amplitude of the optimal control decreases. This reflects the fact that as the system approaches the adiabatic limit, the dynamics under the initial control $\tilde u_0(\tau)$ alone yield a sufficiently high probability of obtaining the ground state.

\begin{figure}[H]
    \centering
    \includegraphics[width=0.31\linewidth]{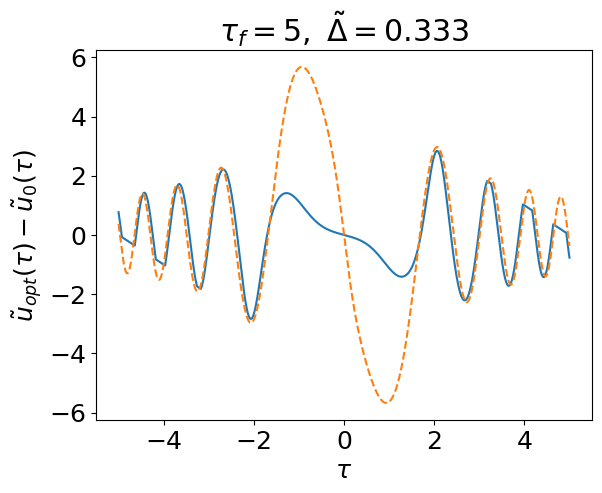}
    \includegraphics[width=0.31\linewidth]{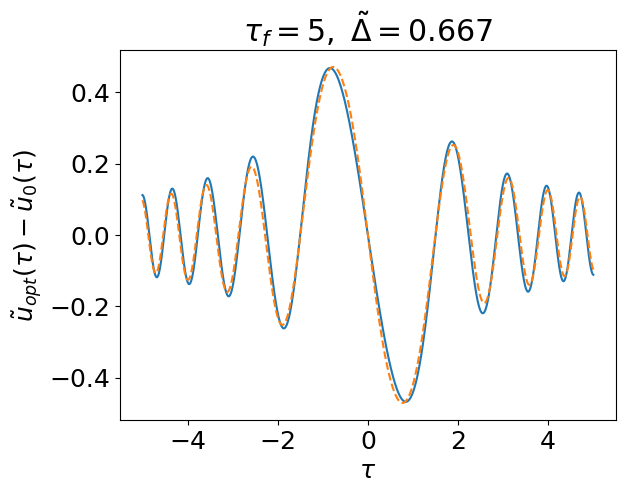}
    \includegraphics[width=0.31\linewidth]{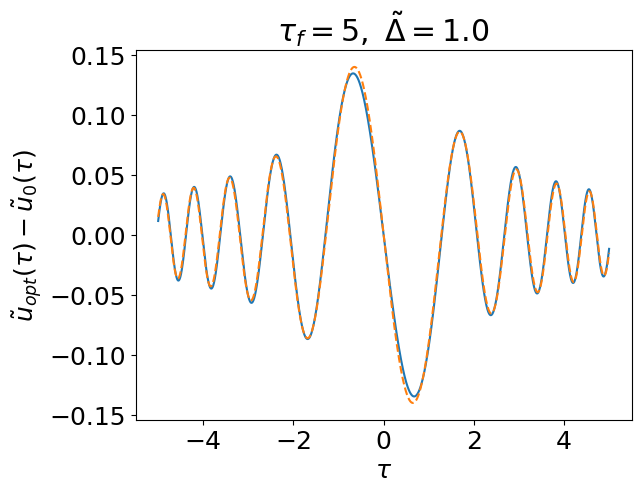}
    \caption{Plots of the difference between the optimal and initial controls for $\tau_f =5, n=1$: (Left) $\tilde \Delta = 2/3$, (Center) $\tilde \Delta = 1$, (Right) $\tilde \Delta = 1/3$. The dashed lines represent the fitting results using $-\tilde{\alpha} \sin(2\int_0^\tau \mathcal{E}_1  (s)ds)/\mathcal{E}_1  (\tau)$, where (Left) $\tilde{\alpha} = 6.25$, (Center) $\tilde{\alpha} = 0.51$, and (Right) $\tilde{\alpha} = 0.17$. It can be observed that as the system approaches the adiabatic limit, $\tilde u(\tau)-\tilde u_0(\tau)$ is well approximated by the fitting function.}
    \label{fig:diff_u_Delta}
\end{figure}

Additionally, the plots of $\tilde u_{\opt}(\tau)-\tilde u_0(\tau)$ for these controls are shown in Fig.~\ref{fig:diff_u_Delta}. The dashed lines in Fig.~\ref{fig:diff_u_Delta} represent the fitting results using
\begin{align}
    \tilde u_{\mathrm{fit}}(\tau)&=-\frac{\tilde{\alpha}\sin(2\int_0^\tau \mathcal{E}_1  (s)ds)}{\mathcal{E}_1  (\tau)}.\label{eq:u_fitting}
\end{align}
It can be observed that in the adiabatic region, $\tilde u(\tau)-\tilde u_0(\tau)$ is well approximated by the fitting function $\tilde u_{\mathrm{fit}}(\tau)$. Notably, $\tilde\alpha$ is always positive, indicating its role in suppressing nonadiabatic transitions at the time when the energy gap is smallest. 

\section{Analysis of Dynamics for Quantum Annealing-type Time Dependence}\label{Sec:Application}

In quantum annealing, the following form is often assumed~\cite{brady2021optimal,brady2021behavior}:
\begin{align}
    H(t)&=\Delta_x (1-u(t))H_x+\Delta_z u(t)H_z,\quad u(0)=0,\ u(T)=1.
\end{align}
We discuss how the analysis changes in cases where both Hamiltonians depend on time. As discussed below, in such cases, the transition amplitude over a finite time interval without oscillations plays a crucial role. 

\begin{figure}[H]
    \centering
    \includegraphics[width=0.31\linewidth]{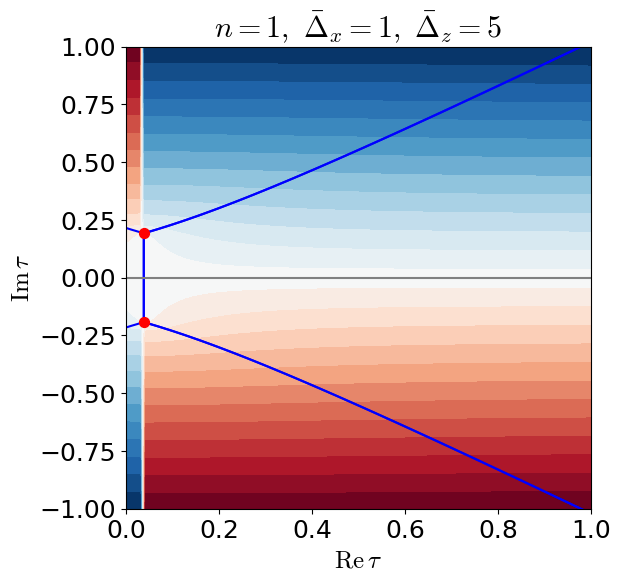}
    \includegraphics[width=0.31\linewidth]{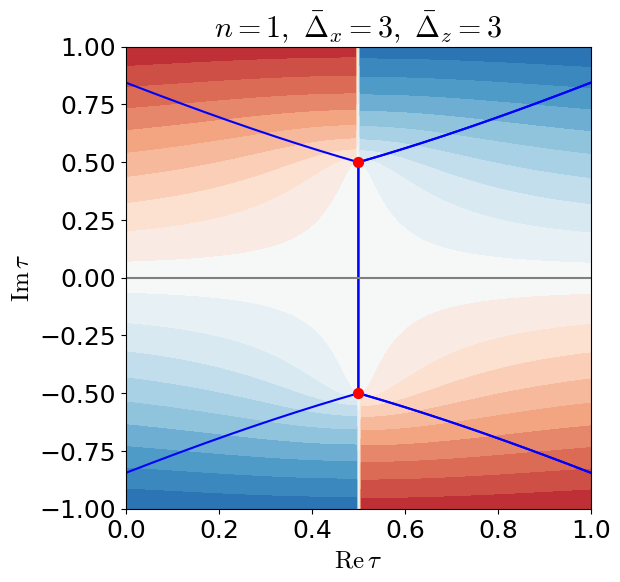}
    \includegraphics[width=0.31\linewidth]{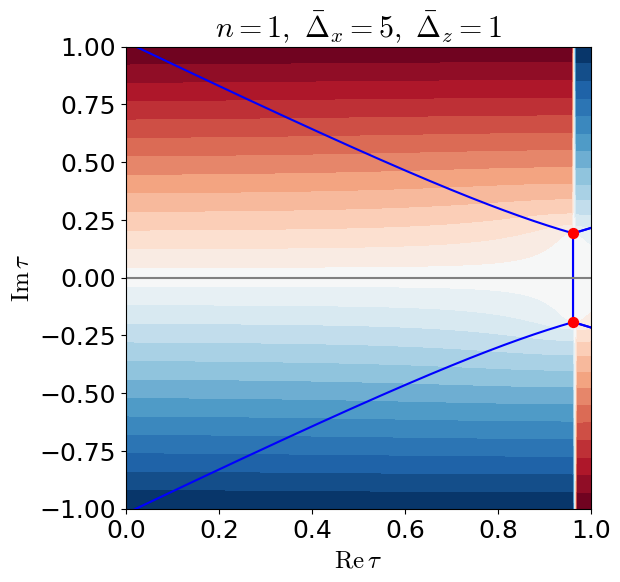}
    \caption{Stokes diagram for the case of $n=1$. In this case, since one Stokes line intersects the real axis within $\tau \in [0,1]$, it can be seen that the number of Stokes lines to be considered is the same as in the case of an infinite time interval.}
    \label{fig:annealin_stokes}
\end{figure}

\begin{figure}[H]
    \centering
    \includegraphics[width=0.31\linewidth]{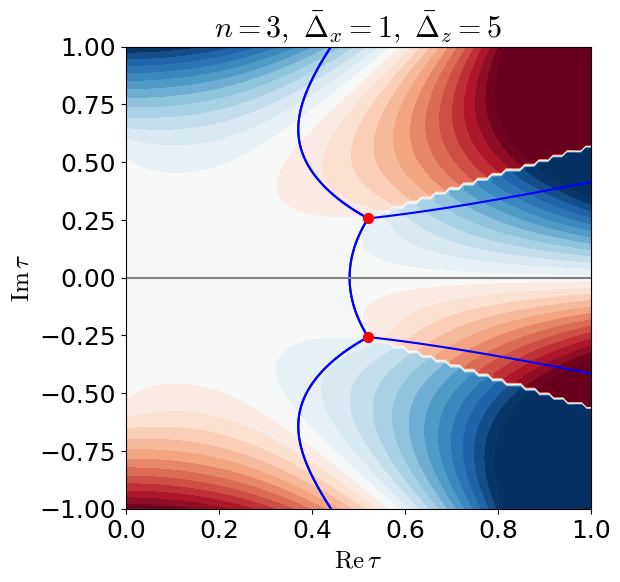}
    \includegraphics[width=0.31\linewidth]{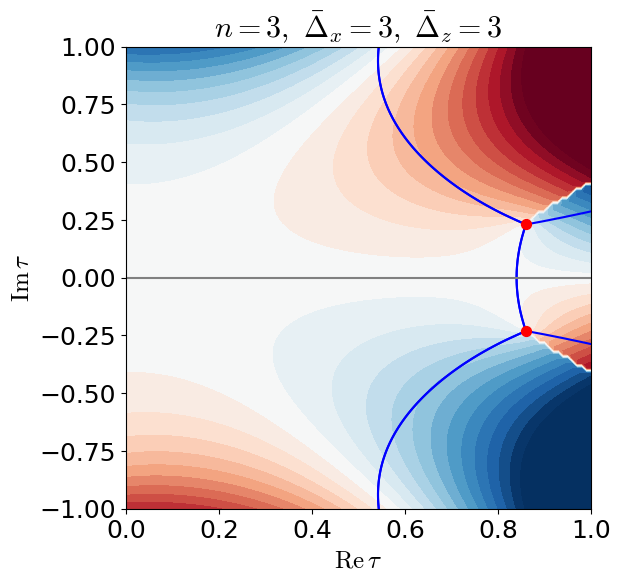}
    \includegraphics[width=0.31\linewidth]{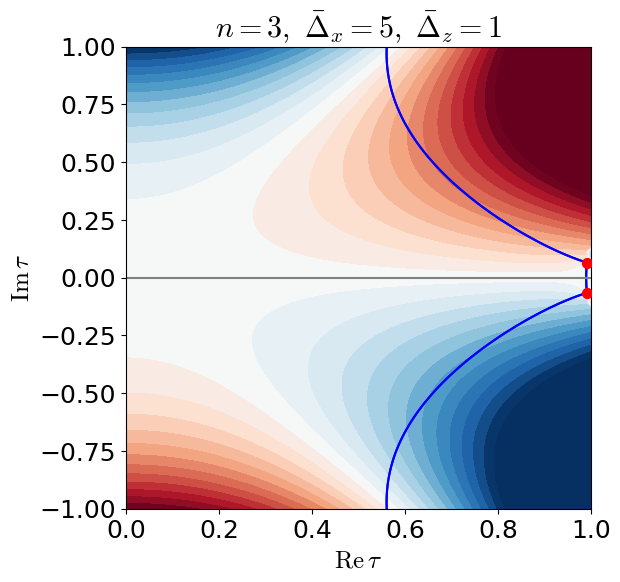}
    \caption{Stokes diagram for the case of $n=3$. In this case, since one Stokes line intersects the real axis within $\tau \in [0,1]$, it can be seen that the number of Stokes lines to be considered has decreased compared to the case of an infinite time interval. Accordingly, $\tilde\alpha_1 = \tilde\alpha_2 = 0$.}
    \label{fig:annealin_stokes_n3}
\end{figure}

Here, we consider the following single qubit Hamiltonian:
\begin{align}
    H(t)&=\Delta_x (1-u(t))\sigma_x+\Delta_z u(t)\sigma_z,\quad u(0)=0,\ u(T)=1.
\end{align}
Assuming $\Delta_x,\Delta_z>0$, the initial state at $t=0$ is $\ket{-}$, and we aim to compute the probability of obtaining $\ket{1}$ at the final time $t=T$. First, we assume the initial control $u_0(t)=(t/T)^n=\tau^n$ and consider the dimensionless Hamiltonian
\begin{align}
    \ms{H}_n(\tau)&=T\Delta_x (1-\tau^n)\sigma_x+T\Delta_z \tau^n\sigma_z=\bar \Delta_x (1-\tau^n)\sigma_x+\bar \Delta_z \tau^n \sigma_z.
\end{align}
We denote the eigenstates of this Hamiltonian as $\ket{\ms{E}_{\pm,n}(\tau)}$. Calculating the positive energy eigenvalue and the turning points, we obtain
\begin{align}
    \ms{E}_n(\tau)&=\sqrt{\qty(1-\tau^n)^2\bar\Delta_x^2+\tau^{2n}\bar\Delta_z^2},\\
    \tau_{c,\pm,k}&=\qty(\frac{\bar\Delta_x\pm i\bar\Delta_z}{\bar\Delta_x^2+\bar\Delta_z^2}\bar\Delta_x)^{1/n}e^{i2\pi k'/n},\quad (k'=0,\cdots,n-1).
\end{align}
Here, the signs $\pm$ and the indices $k$, $k'$ correspond as follows. The sign $+$ corresponds to turning points in the upper half-plane, and the sign $-$ corresponds to turning points in the lower half-plane. Additionally, the index $k$ is taken in order of the real parts, from the largest to the smallest, such that $k = 0, 1, \dots, n-1$. The Stokes diagram for this case is shown in Fig.~\ref{fig:annealin_stokes} and Fig.~\ref{fig:annealin_stokes_n3}. For $n=1$, we find
\begin{align}
    \tau_{r}&=\frac{\bar\Delta_x^2}{\bar\Delta_x^2+\bar\Delta_z^2},
\end{align}
and it can be seen that the number of $\tau_r$ to be considered is the same as in the case of an infinite time interval.
For $n\neq 1$, the Stokes diagram changes compared to the case of infinite time interval. This indicates that only certain Stokes lines need to be considered in the dynamics due to the finite time interval. For example, for $n=3$, as seen in Fig.~\ref{fig:annealin_stokes_n3}, it can be observed that the number of $\tau_{r,k}$ to be considered has decreased to one. Accordingly, $\tilde\alpha_1 = \tilde\alpha_2 = 0$.

\begin{figure}[H]
    \centering
    \includegraphics[width=0.31\linewidth]{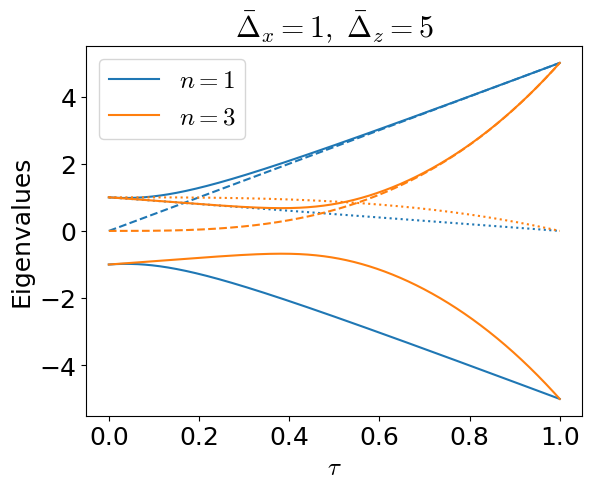}
    \includegraphics[width=0.31\linewidth]{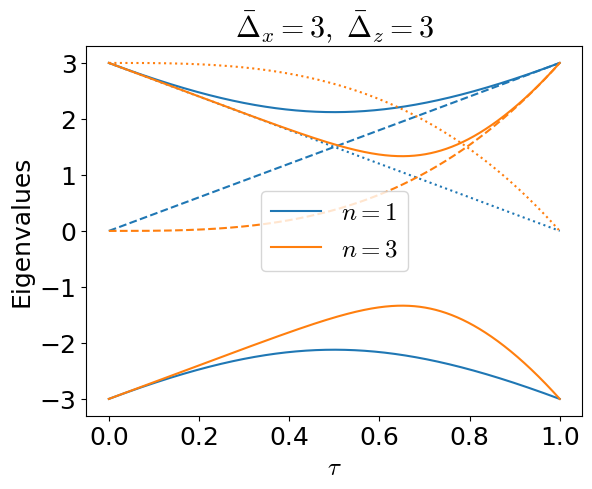}
    \includegraphics[width=0.31\linewidth]{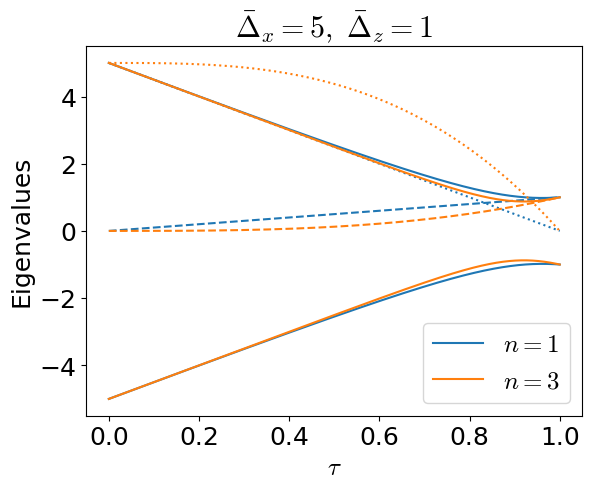}
    \caption{A plot showing the time dependence of the energy. Each graph shows the change in energy levels as a function of $\tau$ for different values of $\Delta_x$ and $\Delta_z$. The dashed and dotted lines represent the time dependence of the coefficients of $\sigma_z$ and $\sigma_x$, respectively. The blue line corresponds to $n=1$ and the orange line corresponds to $n=3$.}
    \label{fig:annealing_energy}
\end{figure}

Next, consider the Hamiltonian with an additional control term:
\begin{align}
    H_n(\tau)&=\ms{H}_n(\tau)+\qty(-\bar\Delta_x \sigma_x+\bar\Delta_z \sigma_z)c_n(\tau) =\ms{H}_n(\tau)+H_c(\tau),
    \\
    c_n(\tau) &= \sum_{k=0}^{n-1}\tilde A_{n,k}(\tau)\sin \qty(\varphi_n(\tau,\tau_{r,k})+\xi_{k})\\,
    \tilde A_{n,k}(t)&=\frac{-\tilde \alpha_k}{\ms{E}_n  (t)}.
\end{align}
Denote the eigenstates of this Hamiltonian as $\ket{E_{n,\pm}(\tau)}$, and let $U_n(\tau)$ represent the time evolution operator under this Hamiltonian, with $\ms{U}_n(\tau)$ representing the evolution operator under $\ms{H}_n(\tau)$. Using perturbative expansion in the Furry picture, we obtain:
\begin{align}
    U_n(\tau)&\simeq \ms{U}_n (\tau)-i\ms{U}_n (\tau)\int_{\tau_0}^\tau ds \ms{U}_n^\dagger(s) H_c(s)\ms{U}_n(s).
\end{align}
The probability amplitude for transitions is then given as:
\begin{align}
    P_e&\simeq \qty| \bra{\ms{E}_{n,+}(\tau_f)}\ms{U}_n (\tau_f)\ket{\ms{E}_{n,-}(\tau_0)} - \frac{\bar\Delta_x\bar\Delta_z}{2}\sum_{k=0}^{n-1} e^{-i\int_{\tau_{r,k}}^{\tau_f} ds' \ms{E}_n(s')}e^{i\int_{\tau_0}^{\tau_{r,k}} ds' \ms{E}_n(s')}e^{-i\xi_{k}}\tilde \alpha_k\int_{\tau_0}^{\tau_f} ds \frac{ 1}{\ms{E}_n^2(s)}|^2.
\end{align}

We consider finite time intervals, so both first and second term are complex. The phase $\xi_k$ is adjusted to ensure the cancellation of differences between the terms. Moreover, not all turning points contribute to the dynamics due to the finite time interval, and the oscillatory components $\tilde\alpha_k$ corresponding to irrelevant turning points are expected to vanish.

\begin{figure}[H]
    \centering
    \includegraphics[width=0.31\linewidth]{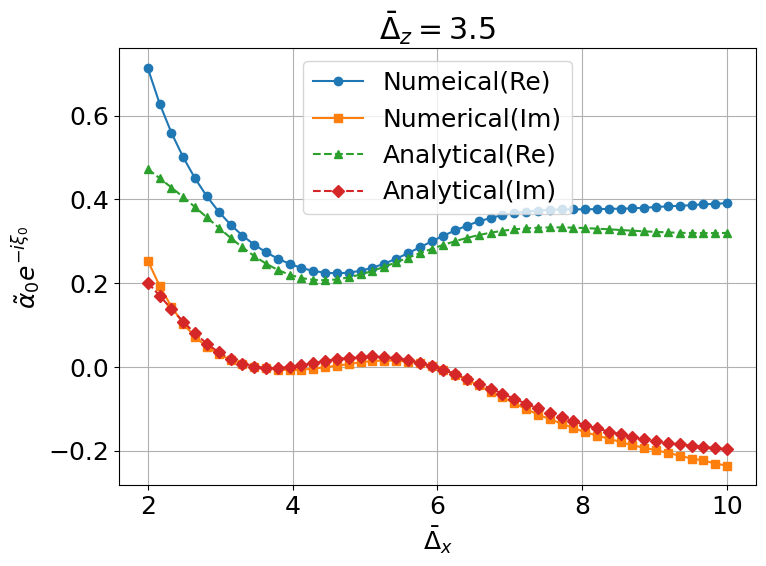}
    \includegraphics[width=0.31\linewidth]{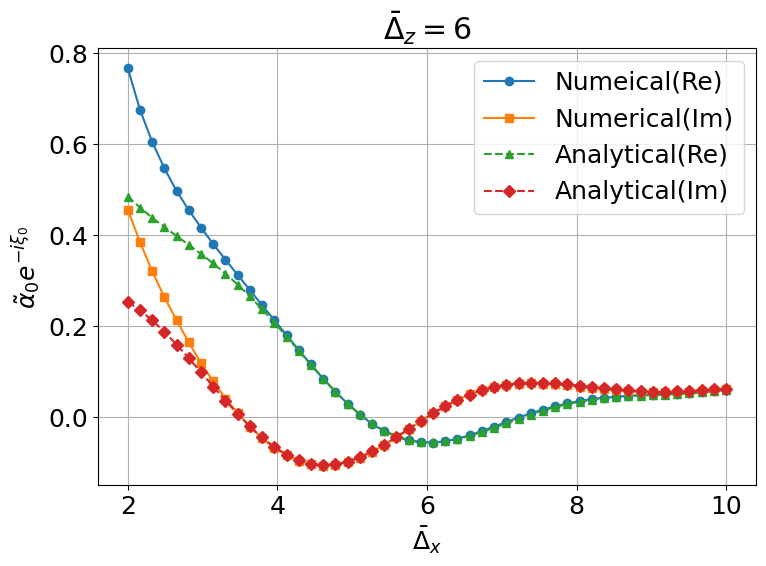}
    \includegraphics[width=0.31\linewidth]{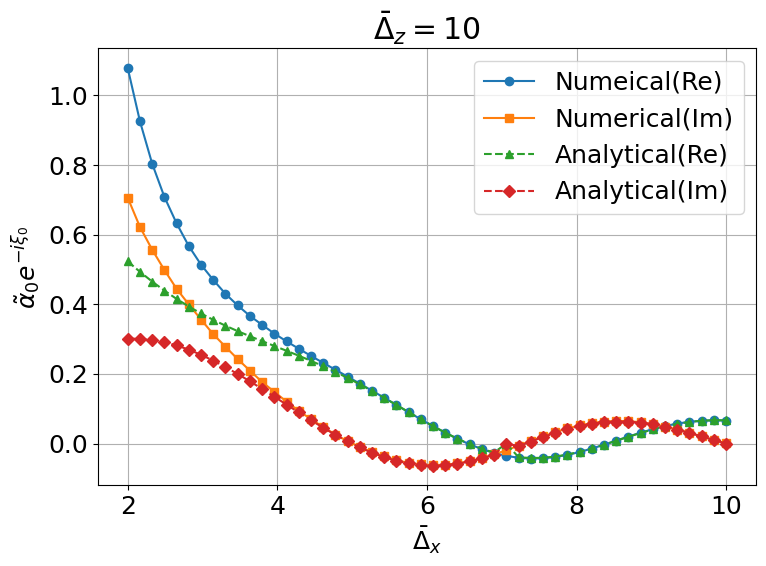}
    \caption{Comparison between (solid) the optimal control obtained by numerical calculation and (dashed) the analytical representation using ``time-dependent resonance" for $n=1$. The real and imaginary parts of $\tilde\alpha_0 e^{-i\xi_0}$ are plotted. As $\bar\Delta_z$ increases, i.e., as we approach the adiabatic limit, it can be seen that the ``time-dependent resonance" control becomes the optimal control.}
    \label{fig:annealing_opt_fit_n1}
\end{figure}

\begin{figure}[H]
    \centering
    \includegraphics[width=0.31\linewidth]{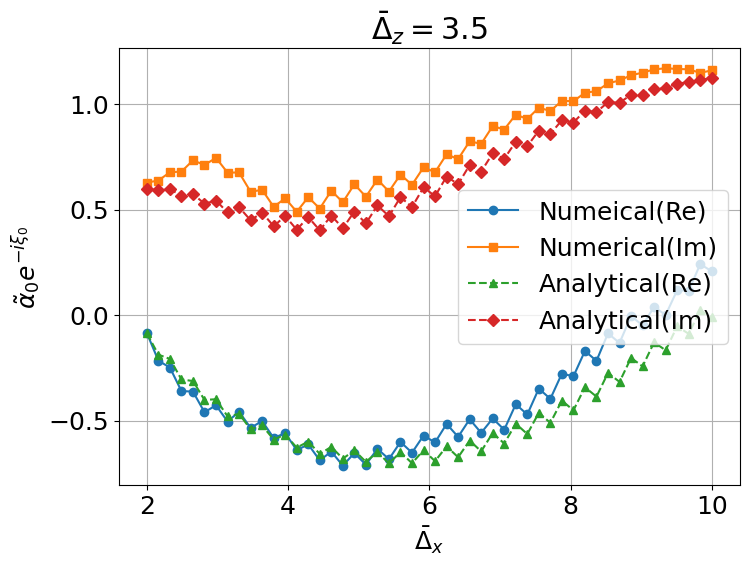}
    \includegraphics[width=0.31\linewidth]{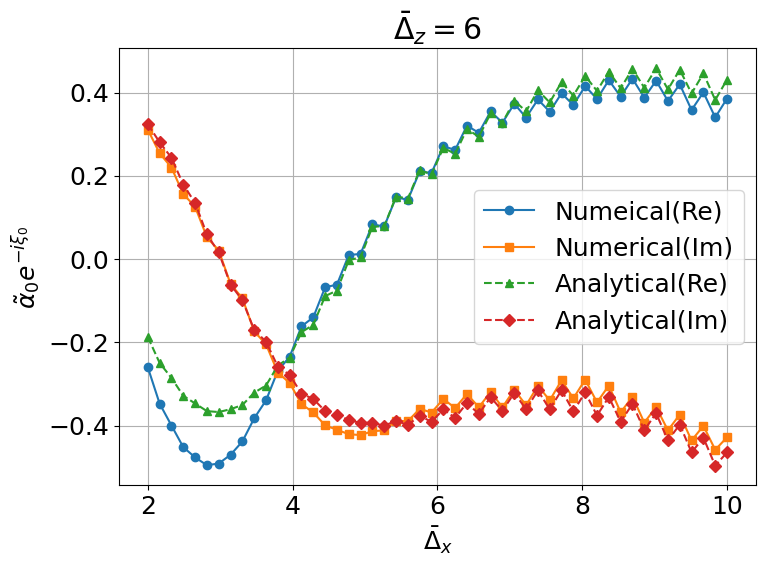}
    \includegraphics[width=0.31\linewidth]{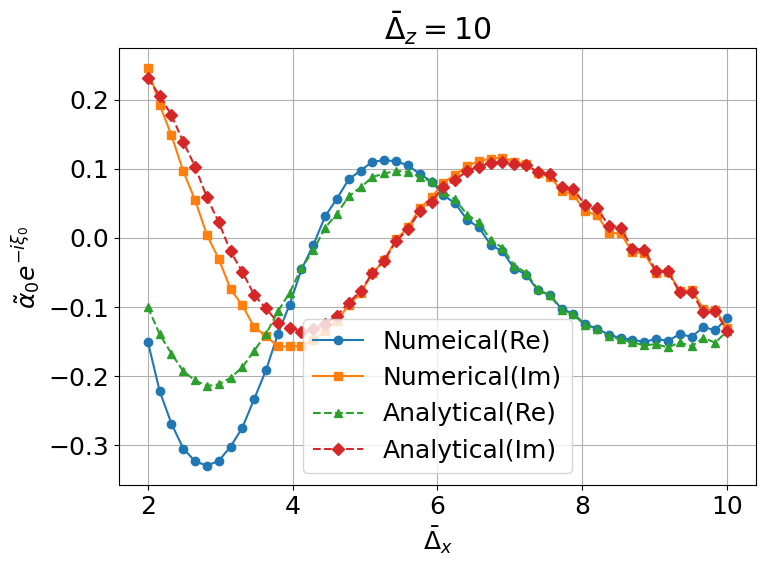}
    \caption{Comparison between (solid) the optimal control obtained by numerical calculation and (dashed) the analytical representation using ``time-dependent" resonance for $n=3$. The real and imaginary parts of $\tilde\alpha_0 e^{-i\xi_0}$ are plotted. As $\bar\Delta_z$ increases, i.e., as we approach the adiabatic limit, it can be seen that the ``time-dependent resonance" control becomes the optimal control.}
    \label{fig:annealing_opt_fit_n3}
\end{figure}

Finally, we verify whether the above considerations hold and whether the chosen control is optimal in the adiabatic regime for $n=1,3$. From Fig.~\ref{fig:annealing_opt_fit_n1} and Fig.~\ref{fig:annealing_opt_fit_n3}, it can be seen that the control using ``time-dependent resonance" in the nearly-adiabatic regime matches the optimal control. Thus, it is found that state control using ``time-dependent resonance" is effective even for optimal control with a finite time interval.

\section{Analysis of Dynamics under ``Time-Dependent Resonance" for Multi-level Systems}

This paper has mainly dealt with two-level systems, but here we will consider how the previously discussed ``time-dependent resonance" can be applied to the control of many-level systems.

We first measure the difference between the ground energy and the excitation energy, and define this function as $f(t)$. If it can be reduced to a two-level system, we have:
\begin{align}
    f(t)^2 &= \Delta_x^2(1-u(t))^2 + \Delta_z^2 u^2(t)
\end{align}
Thus, we can obtain:
\begin{align}
    u(t) &= \frac{\Delta_x^2 \pm \sqrt{\Delta_x^4 - (\Delta_x^2 + \Delta_z^2)(\Delta_x^2 - f^2(t))}}{\Delta_x^2 + \Delta_z^2}
\end{align}

Assuming an initial control of $u_0(t) = t/T = \tau$, we consider the dimensionless Hamiltonian:
\begin{align}
    \ms{H}(\tau) &= T\Delta_x (1-\tau) H_x + T\Delta_z \tau H_\prob = \bar \Delta_x (1-\tau) H_x + \bar \Delta_z \tau H_\prob
\end{align}
Let the eigenstates of this Hamiltonian be denoted as $\ket{\ms{E}_n(\tau)}$. Let the difference between the ground energy and the first excitation energy be denoted as $\delta \ms{E}(\tau)$. For simplicity, assume that the crossing point between the Stokes lines and the real axis for $\delta\ms{E}(\tau)$ is unique, and let this time be denoted as $\tau_r$.

Next, consider the Hamiltonian with the added control term:
\begin{align}
    H(\tau) &= \ms{H}(\tau) + \left( -\bar\Delta_x H_x + \bar\Delta_z H_\prob \right) c(\tau) = \ms{H}(\tau) + H_c(\tau).
\end{align}
Let the eigenstates of this Hamiltonian be denoted as $\ket{E_n(\tau)}$. Let the time evolution operator for this Hamiltonian be denoted as $U(\tau)$, and the time evolution operator for $\ms{H}(\tau)$ be denoted as $\ms{U}(\tau)$. We then have:
\begin{align}
    \bra{E_1(\tau_f)} U(\tau_f) \ket{E_0(\tau_0)} 
    &\simeq \bra{\ms{E}_1(\tau_f)} \ms{U} (\tau_f) \ket{\ms{E}_0(\tau_0)} \\
    &\quad - i \int_{\tau_0}^{\tau_f} ds c(s) e^{-i \int_s^{\tau_f} ds' \ms{E}_1(s')} \bra{\ms{E}_1(s)} \left( -\bar\Delta_x H_x + \bar\Delta_z H_\prob \right) \ket{\ms{E}_0(s)} e^{i \int_{\tau_0}^s \ms{E}_0(s') ds'}.
\end{align}
Here, we define:
\begin{align}
    c(s) &= \frac{-\tilde\alpha}{f_1(s)(\delta\ms{E}(s))^2} \sin \left( \int^s_{\tau_r} \delta\ms{E}(s') ds' + \xi \right) ,\\
    f_1(s) &:= \bra{\ms{E}_1(s)} \left( -\bar\Delta_x H_x + \bar\Delta_z H_\prob \right) \ket{\ms{E}_0(s)}
\end{align}
Then, we can write:
\begin{align}
    \bra{E_1(\tau_f)} U(\tau_f) \ket{E_0(\tau_0)}
    &\simeq \bra{\ms{E}_1(\tau_f)} \ms{U}_n (\tau_f) \ket{\ms{E}_0(\tau_0)} - \frac{1}{2} e^{-i \int_{\tau_r}^{\tau_f} ds' \ms{E}_1(s')} e^{-i \int_{\tau_0}^{\tau_r} ds' \ms{E}_0(s')} e^{-i \xi} \tilde \alpha \int_{\tau_0}^{\tau_f} ds \frac{ 1 }{(\delta\ms{E}(s))^2}
\end{align}
By adjusting $\tilde \alpha$ and $\xi$, it is possible to suppress the transition to the first excited state. However, it is unlikely that this control can suppress transitions to higher energy levels. The optimal control that can be numerically determined can suppress transitions to these higher levels as well, leading to more complex control. As discussed in prior research~\cite{brady2021behavior, brady2021optimal}, it has been numerically confirmed that although the amplitude becomes a complex function, the phase is governed by $\int^s \delta \ms{E}(s') ds'$. This suggests that optimal control in the adiabatic regime may be analytically derived for many-level systems as well.

\bibliography{ref.bib}

\begin{thebibliography}{35}%
\makeatletter
\providecommand \@ifxundefined [1]{%
 \@ifx{#1\undefined}
}%
\providecommand \@ifnum [1]{%
 \ifnum #1\expandafter \@firstoftwo
 \else \expandafter \@secondoftwo
 \fi
}%
\providecommand \@ifx [1]{%
 \ifx #1\expandafter \@firstoftwo
 \else \expandafter \@secondoftwo
 \fi
}%
\providecommand \natexlab [1]{#1}%
\providecommand \enquote  [1]{``#1''}%
\providecommand \bibnamefont  [1]{#1}%
\providecommand \bibfnamefont [1]{#1}%
\providecommand \citenamefont [1]{#1}%
\providecommand \href@noop [0]{\@secondoftwo}%
\providecommand \href [0]{\begingroup \@sanitize@url \@href}%
\providecommand \@href[1]{\@@startlink{#1}\@@href}%
\providecommand \@@href[1]{\endgroup#1\@@endlink}%
\providecommand \@sanitize@url [0]{\catcode `\\12\catcode `\$12\catcode `\&12\catcode `\#12\catcode `\^12\catcode `\_12\catcode `\%12\relax}%
\providecommand \@@startlink[1]{}%
\providecommand \@@endlink[0]{}%
\providecommand \url  [0]{\begingroup\@sanitize@url \@url }%
\providecommand \@url [1]{\endgroup\@href {#1}{\urlprefix }}%
\providecommand \urlprefix  [0]{URL }%
\providecommand \Eprint [0]{\href }%
\providecommand \doibase [0]{https://doi.org/}%
\providecommand \selectlanguage [0]{\@gobble}%
\providecommand \bibinfo  [0]{\@secondoftwo}%
\providecommand \bibfield  [0]{\@secondoftwo}%
\providecommand \translation [1]{[#1]}%
\providecommand \BibitemOpen [0]{}%
\providecommand \bibitemStop [0]{}%
\providecommand \bibitemNoStop [0]{.\EOS\space}%
\providecommand \EOS [0]{\spacefactor3000\relax}%
\providecommand \BibitemShut  [1]{\csname bibitem#1\endcsname}%
\let\auto@bib@innerbib\@empty
\bibitem [{\citenamefont {Werschnik}\ and\ \citenamefont {Gross}(2007)}]{werschnik2007quantum}%
  \BibitemOpen
  \bibfield  {author} {\bibinfo {author} {\bibfnamefont {J.}~\bibnamefont {Werschnik}}\ and\ \bibinfo {author} {\bibfnamefont {E.}~\bibnamefont {Gross}},\ }\bibfield  {title} {\bibinfo {title} {Quantum optimal control theory},\ }\href@noop {} {\bibfield  {journal} {\bibinfo  {journal} {Journal of Physics B: Atomic, Molecular and Optical Physics}\ }\textbf {\bibinfo {volume} {40}},\ \bibinfo {pages} {R175} (\bibinfo {year} {2007})}\BibitemShut {NoStop}%
\bibitem [{\citenamefont {Geng}\ \emph {et~al.}(2016)\citenamefont {Geng}, \citenamefont {Wu}, \citenamefont {Wang}, \citenamefont {Xu}, \citenamefont {Shi}, \citenamefont {Xie}, \citenamefont {Rong},\ and\ \citenamefont {Du}}]{geng2016experimental}%
  \BibitemOpen
  \bibfield  {author} {\bibinfo {author} {\bibfnamefont {J.}~\bibnamefont {Geng}}, \bibinfo {author} {\bibfnamefont {Y.}~\bibnamefont {Wu}}, \bibinfo {author} {\bibfnamefont {X.}~\bibnamefont {Wang}}, \bibinfo {author} {\bibfnamefont {K.}~\bibnamefont {Xu}}, \bibinfo {author} {\bibfnamefont {F.}~\bibnamefont {Shi}}, \bibinfo {author} {\bibfnamefont {Y.}~\bibnamefont {Xie}}, \bibinfo {author} {\bibfnamefont {X.}~\bibnamefont {Rong}},\ and\ \bibinfo {author} {\bibfnamefont {J.}~\bibnamefont {Du}},\ }\bibfield  {title} {\bibinfo {title} {Experimental time-optimal universal control of spin qubits in solids},\ }\href@noop {} {\bibfield  {journal} {\bibinfo  {journal} {Physical review letters}\ }\textbf {\bibinfo {volume} {117}},\ \bibinfo {pages} {170501} (\bibinfo {year} {2016})}\BibitemShut {NoStop}%
\bibitem [{\citenamefont {Rembold}\ \emph {et~al.}(2020)\citenamefont {Rembold}, \citenamefont {Oshnik}, \citenamefont {M{\"u}ller}, \citenamefont {Montangero}, \citenamefont {Calarco},\ and\ \citenamefont {Neu}}]{rembold2020introduction}%
  \BibitemOpen
  \bibfield  {author} {\bibinfo {author} {\bibfnamefont {P.}~\bibnamefont {Rembold}}, \bibinfo {author} {\bibfnamefont {N.}~\bibnamefont {Oshnik}}, \bibinfo {author} {\bibfnamefont {M.~M.}\ \bibnamefont {M{\"u}ller}}, \bibinfo {author} {\bibfnamefont {S.}~\bibnamefont {Montangero}}, \bibinfo {author} {\bibfnamefont {T.}~\bibnamefont {Calarco}},\ and\ \bibinfo {author} {\bibfnamefont {E.}~\bibnamefont {Neu}},\ }\bibfield  {title} {\bibinfo {title} {Introduction to quantum optimal control for quantum sensing with nitrogen-vacancy centers in diamond},\ }\href@noop {} {\bibfield  {journal} {\bibinfo  {journal} {AVS Quantum Science}\ }\textbf {\bibinfo {volume} {2}},\ \bibinfo {pages} {024701} (\bibinfo {year} {2020})}\BibitemShut {NoStop}%
\bibitem [{\citenamefont {Ansel}\ \emph {et~al.}(2024)\citenamefont {Ansel}, \citenamefont {Dionis}, \citenamefont {Arrouas}, \citenamefont {Peaudecerf}, \citenamefont {Gu{\'e}rin}, \citenamefont {Gu{\'e}ry-Odelin},\ and\ \citenamefont {Sugny}}]{ansel2024introduction}%
  \BibitemOpen
  \bibfield  {author} {\bibinfo {author} {\bibfnamefont {Q.}~\bibnamefont {Ansel}}, \bibinfo {author} {\bibfnamefont {E.}~\bibnamefont {Dionis}}, \bibinfo {author} {\bibfnamefont {F.}~\bibnamefont {Arrouas}}, \bibinfo {author} {\bibfnamefont {B.}~\bibnamefont {Peaudecerf}}, \bibinfo {author} {\bibfnamefont {S.}~\bibnamefont {Gu{\'e}rin}}, \bibinfo {author} {\bibfnamefont {D.}~\bibnamefont {Gu{\'e}ry-Odelin}},\ and\ \bibinfo {author} {\bibfnamefont {D.}~\bibnamefont {Sugny}},\ }\bibfield  {title} {\bibinfo {title} {Introduction to theoretical and experimental aspects of quantum optimal control},\ }\href@noop {} {\bibfield  {journal} {\bibinfo  {journal} {Journal of Physics B: Atomic, Molecular and Optical Physics}\ }\textbf {\bibinfo {volume} {57}},\ \bibinfo {pages} {133001} (\bibinfo {year} {2024})}\BibitemShut {NoStop}%
\bibitem [{\citenamefont {Kadowaki}\ and\ \citenamefont {Nishimori}(1998)}]{kadowaki1998quantum}%
  \BibitemOpen
  \bibfield  {author} {\bibinfo {author} {\bibfnamefont {T.}~\bibnamefont {Kadowaki}}\ and\ \bibinfo {author} {\bibfnamefont {H.}~\bibnamefont {Nishimori}},\ }\bibfield  {title} {\bibinfo {title} {Quantum annealing in the transverse ising model},\ }\href@noop {} {\bibfield  {journal} {\bibinfo  {journal} {Physical Review E}\ }\textbf {\bibinfo {volume} {58}},\ \bibinfo {pages} {5355} (\bibinfo {year} {1998})}\BibitemShut {NoStop}%
\bibitem [{\citenamefont {Farhi}\ \emph {et~al.}(2001)\citenamefont {Farhi}, \citenamefont {Goldstone}, \citenamefont {Gutmann}, \citenamefont {Lapan}, \citenamefont {Lundgren},\ and\ \citenamefont {Preda}}]{farhi2001quantum}%
  \BibitemOpen
  \bibfield  {author} {\bibinfo {author} {\bibfnamefont {E.}~\bibnamefont {Farhi}}, \bibinfo {author} {\bibfnamefont {J.}~\bibnamefont {Goldstone}}, \bibinfo {author} {\bibfnamefont {S.}~\bibnamefont {Gutmann}}, \bibinfo {author} {\bibfnamefont {J.}~\bibnamefont {Lapan}}, \bibinfo {author} {\bibfnamefont {A.}~\bibnamefont {Lundgren}},\ and\ \bibinfo {author} {\bibfnamefont {D.}~\bibnamefont {Preda}},\ }\bibfield  {title} {\bibinfo {title} {A quantum adiabatic evolution algorithm applied to random instances of an np-complete problem},\ }\href@noop {} {\bibfield  {journal} {\bibinfo  {journal} {Science}\ }\textbf {\bibinfo {volume} {292}},\ \bibinfo {pages} {472} (\bibinfo {year} {2001})}\BibitemShut {NoStop}%
\bibitem [{\citenamefont {Farhi}\ \emph {et~al.}(2014)\citenamefont {Farhi}, \citenamefont {Goldstone},\ and\ \citenamefont {Gutmann}}]{farhi2014quantum}%
  \BibitemOpen
  \bibfield  {author} {\bibinfo {author} {\bibfnamefont {E.}~\bibnamefont {Farhi}}, \bibinfo {author} {\bibfnamefont {J.}~\bibnamefont {Goldstone}},\ and\ \bibinfo {author} {\bibfnamefont {S.}~\bibnamefont {Gutmann}},\ }\bibfield  {title} {\bibinfo {title} {A quantum approximate optimization algorithm},\ }\href@noop {} {\bibfield  {journal} {\bibinfo  {journal} {arXiv preprint arXiv:1411.4028}\ } (\bibinfo {year} {2014})}\BibitemShut {NoStop}%
\bibitem [{\citenamefont {Yang}\ \emph {et~al.}(2017)\citenamefont {Yang}, \citenamefont {Rahmani}, \citenamefont {Shabani}, \citenamefont {Neven},\ and\ \citenamefont {Chamon}}]{yang2017optimizing}%
  \BibitemOpen
  \bibfield  {author} {\bibinfo {author} {\bibfnamefont {Z.-C.}\ \bibnamefont {Yang}}, \bibinfo {author} {\bibfnamefont {A.}~\bibnamefont {Rahmani}}, \bibinfo {author} {\bibfnamefont {A.}~\bibnamefont {Shabani}}, \bibinfo {author} {\bibfnamefont {H.}~\bibnamefont {Neven}},\ and\ \bibinfo {author} {\bibfnamefont {C.}~\bibnamefont {Chamon}},\ }\bibfield  {title} {\bibinfo {title} {Optimizing variational quantum algorithms using pontryagin’s minimum principle},\ }\href@noop {} {\bibfield  {journal} {\bibinfo  {journal} {Physical Review X}\ }\textbf {\bibinfo {volume} {7}},\ \bibinfo {pages} {021027} (\bibinfo {year} {2017})}\BibitemShut {NoStop}%
\bibitem [{\citenamefont {Brady}\ \emph {et~al.}(2021{\natexlab{a}})\citenamefont {Brady}, \citenamefont {Baldwin}, \citenamefont {Bapat}, \citenamefont {Kharkov},\ and\ \citenamefont {Gorshkov}}]{brady2021optimal}%
  \BibitemOpen
  \bibfield  {author} {\bibinfo {author} {\bibfnamefont {L.~T.}\ \bibnamefont {Brady}}, \bibinfo {author} {\bibfnamefont {C.~L.}\ \bibnamefont {Baldwin}}, \bibinfo {author} {\bibfnamefont {A.}~\bibnamefont {Bapat}}, \bibinfo {author} {\bibfnamefont {Y.}~\bibnamefont {Kharkov}},\ and\ \bibinfo {author} {\bibfnamefont {A.~V.}\ \bibnamefont {Gorshkov}},\ }\bibfield  {title} {\bibinfo {title} {Optimal protocols in quantum annealing and quantum approximate optimization algorithm problems},\ }\href@noop {} {\bibfield  {journal} {\bibinfo  {journal} {Physical Review Letters}\ }\textbf {\bibinfo {volume} {126}},\ \bibinfo {pages} {070505} (\bibinfo {year} {2021}{\natexlab{a}})}\BibitemShut {NoStop}%
\bibitem [{\citenamefont {Gu{\'e}ry-Odelin}\ \emph {et~al.}(2019)\citenamefont {Gu{\'e}ry-Odelin}, \citenamefont {Ruschhaupt}, \citenamefont {Kiely}, \citenamefont {Torrontegui}, \citenamefont {Mart{\'\i}nez-Garaot},\ and\ \citenamefont {Muga}}]{guery2019shortcuts}%
  \BibitemOpen
  \bibfield  {author} {\bibinfo {author} {\bibfnamefont {D.}~\bibnamefont {Gu{\'e}ry-Odelin}}, \bibinfo {author} {\bibfnamefont {A.}~\bibnamefont {Ruschhaupt}}, \bibinfo {author} {\bibfnamefont {A.}~\bibnamefont {Kiely}}, \bibinfo {author} {\bibfnamefont {E.}~\bibnamefont {Torrontegui}}, \bibinfo {author} {\bibfnamefont {S.}~\bibnamefont {Mart{\'\i}nez-Garaot}},\ and\ \bibinfo {author} {\bibfnamefont {J.~G.}\ \bibnamefont {Muga}},\ }\bibfield  {title} {\bibinfo {title} {Shortcuts to adiabaticity: Concepts, methods, and applications},\ }\href@noop {} {\bibfield  {journal} {\bibinfo  {journal} {Reviews of Modern Physics}\ }\textbf {\bibinfo {volume} {91}},\ \bibinfo {pages} {045001} (\bibinfo {year} {2019})}\BibitemShut {NoStop}%
\bibitem [{\citenamefont {Caneva}\ \emph {et~al.}(2009)\citenamefont {Caneva}, \citenamefont {Murphy}, \citenamefont {Calarco}, \citenamefont {Fazio}, \citenamefont {Montangero}, \citenamefont {Giovannetti},\ and\ \citenamefont {Santoro}}]{caneva2009optimal}%
  \BibitemOpen
  \bibfield  {author} {\bibinfo {author} {\bibfnamefont {T.}~\bibnamefont {Caneva}}, \bibinfo {author} {\bibfnamefont {M.}~\bibnamefont {Murphy}}, \bibinfo {author} {\bibfnamefont {T.}~\bibnamefont {Calarco}}, \bibinfo {author} {\bibfnamefont {R.}~\bibnamefont {Fazio}}, \bibinfo {author} {\bibfnamefont {S.}~\bibnamefont {Montangero}}, \bibinfo {author} {\bibfnamefont {V.}~\bibnamefont {Giovannetti}},\ and\ \bibinfo {author} {\bibfnamefont {G.~E.}\ \bibnamefont {Santoro}},\ }\bibfield  {title} {\bibinfo {title} {Optimal control at the quantum speed limit},\ }\href@noop {} {\bibfield  {journal} {\bibinfo  {journal} {Physical review letters}\ }\textbf {\bibinfo {volume} {103}},\ \bibinfo {pages} {240501} (\bibinfo {year} {2009})}\BibitemShut {NoStop}%
\bibitem [{\citenamefont {Hu}\ \emph {et~al.}(2019)\citenamefont {Hu}, \citenamefont {Santos}, \citenamefont {Cui}, \citenamefont {Huang}, \citenamefont {Sarandy}, \citenamefont {Li},\ and\ \citenamefont {Guo}}]{hu2019adiabatic}%
  \BibitemOpen
  \bibfield  {author} {\bibinfo {author} {\bibfnamefont {C.-K.}\ \bibnamefont {Hu}}, \bibinfo {author} {\bibfnamefont {A.~C.}\ \bibnamefont {Santos}}, \bibinfo {author} {\bibfnamefont {J.-M.}\ \bibnamefont {Cui}}, \bibinfo {author} {\bibfnamefont {Y.-F.}\ \bibnamefont {Huang}}, \bibinfo {author} {\bibfnamefont {M.~S.}\ \bibnamefont {Sarandy}}, \bibinfo {author} {\bibfnamefont {C.-F.}\ \bibnamefont {Li}},\ and\ \bibinfo {author} {\bibfnamefont {G.-C.}\ \bibnamefont {Guo}},\ }\bibfield  {title} {\bibinfo {title} {Adiabatic quantum dynamics under decoherence in a controllable trapped-ion setup},\ }\href@noop {} {\bibfield  {journal} {\bibinfo  {journal} {Physical Review A}\ }\textbf {\bibinfo {volume} {99}},\ \bibinfo {pages} {062320} (\bibinfo {year} {2019})}\BibitemShut {NoStop}%
\bibitem [{\citenamefont {Kandel}\ \emph {et~al.}(2021)\citenamefont {Kandel}, \citenamefont {Qiao}, \citenamefont {Fallahi}, \citenamefont {Gardner}, \citenamefont {Manfra},\ and\ \citenamefont {Nichol}}]{kandel2021adiabatic}%
  \BibitemOpen
  \bibfield  {author} {\bibinfo {author} {\bibfnamefont {Y.~P.}\ \bibnamefont {Kandel}}, \bibinfo {author} {\bibfnamefont {H.}~\bibnamefont {Qiao}}, \bibinfo {author} {\bibfnamefont {S.}~\bibnamefont {Fallahi}}, \bibinfo {author} {\bibfnamefont {G.~C.}\ \bibnamefont {Gardner}}, \bibinfo {author} {\bibfnamefont {M.~J.}\ \bibnamefont {Manfra}},\ and\ \bibinfo {author} {\bibfnamefont {J.~M.}\ \bibnamefont {Nichol}},\ }\bibfield  {title} {\bibinfo {title} {Adiabatic quantum state transfer in a semiconductor quantum-dot spin chain},\ }\href@noop {} {\bibfield  {journal} {\bibinfo  {journal} {Nature communications}\ }\textbf {\bibinfo {volume} {12}},\ \bibinfo {pages} {2156} (\bibinfo {year} {2021})}\BibitemShut {NoStop}%
\bibitem [{\citenamefont {Kr{\'a}l}\ \emph {et~al.}(2007)\citenamefont {Kr{\'a}l}, \citenamefont {Thanopulos},\ and\ \citenamefont {Shapiro}}]{kral2007colloquium}%
  \BibitemOpen
  \bibfield  {author} {\bibinfo {author} {\bibfnamefont {P.}~\bibnamefont {Kr{\'a}l}}, \bibinfo {author} {\bibfnamefont {I.}~\bibnamefont {Thanopulos}},\ and\ \bibinfo {author} {\bibfnamefont {M.}~\bibnamefont {Shapiro}},\ }\bibfield  {title} {\bibinfo {title} {Colloquium: Coherently controlled adiabatic passage},\ }\href@noop {} {\bibfield  {journal} {\bibinfo  {journal} {Reviews of modern physics}\ }\textbf {\bibinfo {volume} {79}},\ \bibinfo {pages} {53} (\bibinfo {year} {2007})}\BibitemShut {NoStop}%
\bibitem [{\citenamefont {Costa}\ \emph {et~al.}(2022)\citenamefont {Costa}, \citenamefont {An}, \citenamefont {Sanders}, \citenamefont {Su}, \citenamefont {Babbush},\ and\ \citenamefont {Berry}}]{costa2022optimal}%
  \BibitemOpen
  \bibfield  {author} {\bibinfo {author} {\bibfnamefont {P.~C.}\ \bibnamefont {Costa}}, \bibinfo {author} {\bibfnamefont {D.}~\bibnamefont {An}}, \bibinfo {author} {\bibfnamefont {Y.~R.}\ \bibnamefont {Sanders}}, \bibinfo {author} {\bibfnamefont {Y.}~\bibnamefont {Su}}, \bibinfo {author} {\bibfnamefont {R.}~\bibnamefont {Babbush}},\ and\ \bibinfo {author} {\bibfnamefont {D.~W.}\ \bibnamefont {Berry}},\ }\bibfield  {title} {\bibinfo {title} {Optimal scaling quantum linear-systems solver via discrete adiabatic theorem},\ }\href@noop {} {\bibfield  {journal} {\bibinfo  {journal} {PRX quantum}\ }\textbf {\bibinfo {volume} {3}},\ \bibinfo {pages} {040303} (\bibinfo {year} {2022})}\BibitemShut {NoStop}%
\bibitem [{\citenamefont {Kovalsky}\ \emph {et~al.}(2023)\citenamefont {Kovalsky}, \citenamefont {Calderon-Vargas}, \citenamefont {Grace}, \citenamefont {Magann}, \citenamefont {Larsen}, \citenamefont {Baczewski},\ and\ \citenamefont {Sarovar}}]{kovalsky2023self}%
  \BibitemOpen
  \bibfield  {author} {\bibinfo {author} {\bibfnamefont {L.~K.}\ \bibnamefont {Kovalsky}}, \bibinfo {author} {\bibfnamefont {F.~A.}\ \bibnamefont {Calderon-Vargas}}, \bibinfo {author} {\bibfnamefont {M.~D.}\ \bibnamefont {Grace}}, \bibinfo {author} {\bibfnamefont {A.~B.}\ \bibnamefont {Magann}}, \bibinfo {author} {\bibfnamefont {J.~B.}\ \bibnamefont {Larsen}}, \bibinfo {author} {\bibfnamefont {A.~D.}\ \bibnamefont {Baczewski}},\ and\ \bibinfo {author} {\bibfnamefont {M.}~\bibnamefont {Sarovar}},\ }\bibfield  {title} {\bibinfo {title} {Self-healing of trotter error in digital adiabatic state preparation},\ }\href@noop {} {\bibfield  {journal} {\bibinfo  {journal} {Physical Review Letters}\ }\textbf {\bibinfo {volume} {131}},\ \bibinfo {pages} {060602} (\bibinfo {year} {2023})}\BibitemShut {NoStop}%
\bibitem [{\citenamefont {Xu}\ \emph {et~al.}(2017)\citenamefont {Xu}, \citenamefont {Xie}, \citenamefont {Li}, \citenamefont {Xu}, \citenamefont {Wang}, \citenamefont {Ye}, \citenamefont {Kong}, \citenamefont {Geng}, \citenamefont {Duan}, \citenamefont {Shi} \emph {et~al.}}]{xu2017experimental}%
  \BibitemOpen
  \bibfield  {author} {\bibinfo {author} {\bibfnamefont {K.}~\bibnamefont {Xu}}, \bibinfo {author} {\bibfnamefont {T.}~\bibnamefont {Xie}}, \bibinfo {author} {\bibfnamefont {Z.}~\bibnamefont {Li}}, \bibinfo {author} {\bibfnamefont {X.}~\bibnamefont {Xu}}, \bibinfo {author} {\bibfnamefont {M.}~\bibnamefont {Wang}}, \bibinfo {author} {\bibfnamefont {X.}~\bibnamefont {Ye}}, \bibinfo {author} {\bibfnamefont {F.}~\bibnamefont {Kong}}, \bibinfo {author} {\bibfnamefont {J.}~\bibnamefont {Geng}}, \bibinfo {author} {\bibfnamefont {C.}~\bibnamefont {Duan}}, \bibinfo {author} {\bibfnamefont {F.}~\bibnamefont {Shi}}, \emph {et~al.},\ }\bibfield  {title} {\bibinfo {title} {Experimental adiabatic quantum factorization under ambient conditions based on a solid-state single spin system},\ }\href@noop {} {\bibfield  {journal} {\bibinfo  {journal} {Physical review letters}\ }\textbf {\bibinfo {volume} {118}},\ \bibinfo {pages} {130504} (\bibinfo {year} {2017})}\BibitemShut {NoStop}%
\bibitem [{\citenamefont {Albash}\ and\ \citenamefont {Lidar}(2018)}]{albash2018adiabatic}%
  \BibitemOpen
  \bibfield  {author} {\bibinfo {author} {\bibfnamefont {T.}~\bibnamefont {Albash}}\ and\ \bibinfo {author} {\bibfnamefont {D.~A.}\ \bibnamefont {Lidar}},\ }\bibfield  {title} {\bibinfo {title} {Adiabatic quantum computation},\ }\href@noop {} {\bibfield  {journal} {\bibinfo  {journal} {Reviews of Modern Physics}\ }\textbf {\bibinfo {volume} {90}},\ \bibinfo {pages} {015002} (\bibinfo {year} {2018})}\BibitemShut {NoStop}%
\bibitem [{\citenamefont {Wen}\ \emph {et~al.}(2019)\citenamefont {Wen}, \citenamefont {Kong}, \citenamefont {Wei}, \citenamefont {Wang}, \citenamefont {Xin},\ and\ \citenamefont {Long}}]{wen2019experimental}%
  \BibitemOpen
  \bibfield  {author} {\bibinfo {author} {\bibfnamefont {J.}~\bibnamefont {Wen}}, \bibinfo {author} {\bibfnamefont {X.}~\bibnamefont {Kong}}, \bibinfo {author} {\bibfnamefont {S.}~\bibnamefont {Wei}}, \bibinfo {author} {\bibfnamefont {B.}~\bibnamefont {Wang}}, \bibinfo {author} {\bibfnamefont {T.}~\bibnamefont {Xin}},\ and\ \bibinfo {author} {\bibfnamefont {G.}~\bibnamefont {Long}},\ }\bibfield  {title} {\bibinfo {title} {Experimental realization of quantum algorithms for a linear system inspired by adiabatic quantum computing},\ }\href@noop {} {\bibfield  {journal} {\bibinfo  {journal} {Physical Review A}\ }\textbf {\bibinfo {volume} {99}},\ \bibinfo {pages} {012320} (\bibinfo {year} {2019})}\BibitemShut {NoStop}%
\bibitem [{\citenamefont {Brady}\ \emph {et~al.}(2021{\natexlab{b}})\citenamefont {Brady}, \citenamefont {Kocia}, \citenamefont {Bienias}, \citenamefont {Bapat}, \citenamefont {Kharkov},\ and\ \citenamefont {Gorshkov}}]{brady2021behavior}%
  \BibitemOpen
  \bibfield  {author} {\bibinfo {author} {\bibfnamefont {L.~T.}\ \bibnamefont {Brady}}, \bibinfo {author} {\bibfnamefont {L.}~\bibnamefont {Kocia}}, \bibinfo {author} {\bibfnamefont {P.}~\bibnamefont {Bienias}}, \bibinfo {author} {\bibfnamefont {A.}~\bibnamefont {Bapat}}, \bibinfo {author} {\bibfnamefont {Y.}~\bibnamefont {Kharkov}},\ and\ \bibinfo {author} {\bibfnamefont {A.~V.}\ \bibnamefont {Gorshkov}},\ }\bibfield  {title} {\bibinfo {title} {Behavior of analog quantum algorithms},\ }\href@noop {} {\bibfield  {journal} {\bibinfo  {journal} {arXiv preprint arXiv:2107.01218}\ } (\bibinfo {year} {2021}{\natexlab{b}})}\BibitemShut {NoStop}%
\bibitem [{\citenamefont {Landau}(1932)}]{landau1932theorie}%
  \BibitemOpen
  \bibfield  {author} {\bibinfo {author} {\bibfnamefont {L.}~\bibnamefont {Landau}},\ }\bibfield  {title} {\bibinfo {title} {On the theory of transfer of energy at collisions ii},\ }\href@noop {} {\bibfield  {journal} {\bibinfo  {journal} {Z. Sowjetunion}\ }\textbf {\bibinfo {volume} {2}},\ \bibinfo {pages} {46} (\bibinfo {year} {1932})}\BibitemShut {NoStop}%
\bibitem [{\citenamefont {Zener}(1932)}]{zener1932non}%
  \BibitemOpen
  \bibfield  {author} {\bibinfo {author} {\bibfnamefont {C.}~\bibnamefont {Zener}},\ }\bibfield  {title} {\bibinfo {title} {Non-adiabatic crossing of energy levels},\ }\href {https://doi.org/10.1098/rspa.1932.0165} {\bibfield  {journal} {\bibinfo  {journal} {Proc. R. Soc. A}\ }\textbf {\bibinfo {volume} {137}},\ \bibinfo {pages} {696} (\bibinfo {year} {1932})}\BibitemShut {NoStop}%
\bibitem [{\citenamefont {St{\"u}ckelberg}(1932)}]{stueckelberg1932two}%
  \BibitemOpen
  \bibfield  {author} {\bibinfo {author} {\bibfnamefont {E.~C.~G.}\ \bibnamefont {St{\"u}ckelberg}},\ }\bibfield  {title} {\bibinfo {title} {Theory of inelastic collisions between atoms(theory of inelastic collisions between atoms, using two simultaneous differential equations)},\ }\href@noop {} {\bibfield  {journal} {\bibinfo  {journal} {Helv. Phys. Acta}\ }\textbf {\bibinfo {volume} {5}},\ \bibinfo {pages} {369} (\bibinfo {year} {1932})}\BibitemShut {NoStop}%
\bibitem [{\citenamefont {Majorana}(1932)}]{majorana1932atoms}%
  \BibitemOpen
  \bibfield  {author} {\bibinfo {author} {\bibfnamefont {E.}~\bibnamefont {Majorana}},\ }\bibfield  {title} {\bibinfo {title} {Atomi orientati in campo magnetico variabile},\ }\href@noop {} {\bibfield  {journal} {\bibinfo  {journal} {Nuovo Cimento}\ }\textbf {\bibinfo {volume} {9}},\ \bibinfo {pages} {43} (\bibinfo {year} {1932})}\BibitemShut {NoStop}%
\bibitem [{\citenamefont {Vitanov}\ and\ \citenamefont {Suominen}(1999)}]{vitanov1999nonlinear}%
  \BibitemOpen
  \bibfield  {author} {\bibinfo {author} {\bibfnamefont {N.}~\bibnamefont {Vitanov}}\ and\ \bibinfo {author} {\bibfnamefont {K.-A.}\ \bibnamefont {Suominen}},\ }\bibfield  {title} {\bibinfo {title} {Nonlinear level-crossing models},\ }\href@noop {} {\bibfield  {journal} {\bibinfo  {journal} {Physical Review A}\ }\textbf {\bibinfo {volume} {59}},\ \bibinfo {pages} {4580} (\bibinfo {year} {1999})}\BibitemShut {NoStop}%
\bibitem [{\citenamefont {Lehto}\ and\ \citenamefont {Suominen}(2012)}]{lehto2012superparabolic}%
  \BibitemOpen
  \bibfield  {author} {\bibinfo {author} {\bibfnamefont {J.}~\bibnamefont {Lehto}}\ and\ \bibinfo {author} {\bibfnamefont {K.-A.}\ \bibnamefont {Suominen}},\ }\bibfield  {title} {\bibinfo {title} {Superparabolic level-glancing models for two-state quantum systems},\ }\href@noop {} {\bibfield  {journal} {\bibinfo  {journal} {Physical Review A―Atomic, Molecular, and Optical Physics}\ }\textbf {\bibinfo {volume} {86}},\ \bibinfo {pages} {033415} (\bibinfo {year} {2012})}\BibitemShut {NoStop}%
\bibitem [{\citenamefont {Dykhne}(1962)}]{dykhne1962adiabatic}%
  \BibitemOpen
  \bibfield  {author} {\bibinfo {author} {\bibfnamefont {A.~M.}\ \bibnamefont {Dykhne}},\ }\bibfield  {title} {\bibinfo {title} {Adiabatic perturbation of discrete spectrum states},\ }\href@noop {} {\bibfield  {journal} {\bibinfo  {journal} {Sov. Phys. JETP}\ }\textbf {\bibinfo {volume} {14}},\ \bibinfo {pages} {1} (\bibinfo {year} {1962})}\BibitemShut {NoStop}%
\bibitem [{\citenamefont {Davis}\ and\ \citenamefont {Pechukas}(1976)}]{davis1976nonadiabatic}%
  \BibitemOpen
  \bibfield  {author} {\bibinfo {author} {\bibfnamefont {J.~P.}\ \bibnamefont {Davis}}\ and\ \bibinfo {author} {\bibfnamefont {P.}~\bibnamefont {Pechukas}},\ }\bibfield  {title} {\bibinfo {title} {{Nonadiabatic transitions induced by a time‐dependent Hamiltonian in the semiclassical/adiabatic limit: The two‐state case}},\ }\href {https://doi.org/10.1063/1.432648} {\bibfield  {journal} {\bibinfo  {journal} {J. Chem. Phys.}\ }\textbf {\bibinfo {volume} {64}},\ \bibinfo {pages} {3129} (\bibinfo {year} {1976})}\BibitemShut {NoStop}%
\bibitem [{sup()}]{suppl_ref}%
  \BibitemOpen
  \href@noop {} {}\bibinfo {note} {See Supplemental Material. The Supplemental Material has four sections: (i) a brief introduction to the exact WKB analysis; (ii) detailed information on the numerical calculations for optimal control; (iii) an analysis of the dynamics for quantum annealing-type time dependence; (iv) the potential for extension to multi-level systems.}\BibitemShut {Stop}%
\bibitem [{\citenamefont {Shevchenko}\ \emph {et~al.}(2010)\citenamefont {Shevchenko}, \citenamefont {Ashhab},\ and\ \citenamefont {Nori}}]{shevchenko2010landau}%
  \BibitemOpen
  \bibfield  {author} {\bibinfo {author} {\bibfnamefont {S.~N.}\ \bibnamefont {Shevchenko}}, \bibinfo {author} {\bibfnamefont {S.}~\bibnamefont {Ashhab}},\ and\ \bibinfo {author} {\bibfnamefont {F.}~\bibnamefont {Nori}},\ }\bibfield  {title} {\bibinfo {title} {Landau--zener--st{\"u}ckelberg interferometry},\ }\href@noop {} {\bibfield  {journal} {\bibinfo  {journal} {Physics Reports}\ }\textbf {\bibinfo {volume} {492}},\ \bibinfo {pages} {1} (\bibinfo {year} {2010})}\BibitemShut {NoStop}%
\bibitem [{\citenamefont {Suzuki}\ and\ \citenamefont {Nakazato}(2022)}]{suzuki2022generalized}%
  \BibitemOpen
  \bibfield  {author} {\bibinfo {author} {\bibfnamefont {T.}~\bibnamefont {Suzuki}}\ and\ \bibinfo {author} {\bibfnamefont {H.}~\bibnamefont {Nakazato}},\ }\bibfield  {title} {\bibinfo {title} {Generalized adiabatic impulse approximation},\ }\href@noop {} {\bibfield  {journal} {\bibinfo  {journal} {Physical Review A}\ }\textbf {\bibinfo {volume} {105}},\ \bibinfo {pages} {022211} (\bibinfo {year} {2022})}\BibitemShut {NoStop}%
\bibitem [{\citenamefont {Furry}(1951)}]{furry1951bound}%
  \BibitemOpen
  \bibfield  {author} {\bibinfo {author} {\bibfnamefont {W.}~\bibnamefont {Furry}},\ }\bibfield  {title} {\bibinfo {title} {On bound states and scattering in positron theory},\ }\href@noop {} {\bibfield  {journal} {\bibinfo  {journal} {Physical Review}\ }\textbf {\bibinfo {volume} {81}},\ \bibinfo {pages} {115} (\bibinfo {year} {1951})}\BibitemShut {NoStop}%
\bibitem [{\citenamefont {Suzuki}\ and\ \citenamefont {Iwamura}(2023)}]{suzuki2023kibble}%
  \BibitemOpen
  \bibfield  {author} {\bibinfo {author} {\bibfnamefont {T.}~\bibnamefont {Suzuki}}\ and\ \bibinfo {author} {\bibfnamefont {K.}~\bibnamefont {Iwamura}},\ }\bibfield  {title} {\bibinfo {title} {Kibble-zurek scaling in the quantum ising chain with a time-periodic perturbation},\ }\href@noop {} {\bibfield  {journal} {\bibinfo  {journal} {Physical Review B}\ }\textbf {\bibinfo {volume} {108}},\ \bibinfo {pages} {014110} (\bibinfo {year} {2023})}\BibitemShut {NoStop}%
\bibitem [{\citenamefont {Taya}(2020)}]{taya2020dynamically}%
  \BibitemOpen
  \bibfield  {author} {\bibinfo {author} {\bibfnamefont {H.}~\bibnamefont {Taya}},\ }\bibfield  {title} {\bibinfo {title} {Dynamically assisted schwinger mechanism and chirality production in parallel electromagnetic field},\ }\href@noop {} {\bibfield  {journal} {\bibinfo  {journal} {Physical Review Research}\ }\textbf {\bibinfo {volume} {2}},\ \bibinfo {pages} {023257} (\bibinfo {year} {2020})}\BibitemShut {NoStop}%
\bibitem [{\citenamefont {Suzuki}\ \emph {et~al.}(2024)\citenamefont {Suzuki}, \citenamefont {Taniguchi},\ and\ \citenamefont {Iwamura}}]{suzuki2024exact}%
  \BibitemOpen
  \bibfield  {author} {\bibinfo {author} {\bibfnamefont {T.}~\bibnamefont {Suzuki}}, \bibinfo {author} {\bibfnamefont {E.}~\bibnamefont {Taniguchi}},\ and\ \bibinfo {author} {\bibfnamefont {K.}~\bibnamefont {Iwamura}},\ }\bibfield  {title} {\bibinfo {title} {Exact wkb analysis for adiabatic discrete-level hamiltonians},\ }\href@noop {} {\bibfield  {journal} {\bibinfo  {journal} {Physical Review A}\ }\textbf {\bibinfo {volume} {109}},\ \bibinfo {pages} {022225} (\bibinfo {year} {2024})}\BibitemShut {NoStop}%
\end{thebibliography}%

\end{document}